\documentclass[12pt]{article}
\pdfoutput=1

\usepackage{authblk}
\usepackage[utf8]{inputenc}
\usepackage[T1]{fontenc}
\usepackage{float}
\usepackage{authblk}
\usepackage[normalem]{ulem}
\usepackage{amsmath}
\usepackage{amsfonts}
\usepackage{amsthm}
\usepackage{adjustbox}
\usepackage{graphicx}
\usepackage{booktabs}
\usepackage{mathtools}
\usepackage{fullpage}
\usepackage[mathlines]{lineno}
\usepackage{subcaption}
\usepackage[style=numeric-comp,sorting=none,maxnames=25]{biblatex} 
\usepackage{hanging}
\usepackage{longtable}
\usepackage{algorithm}
\usepackage{algorithmic}
\usepackage{comment}

\usepackage[utf8]{inputenc}
\usepackage{booktabs} 
\usepackage{caption} 

\usepackage{multirow}
\usepackage[table,xcdraw]{xcolor}
\usepackage[colorlinks = true,
            linkcolor = black,
            urlcolor  = black,
            citecolor = teal]{hyperref}

\title{Active Inference and Intentional Behaviour}

\author[1,2]{Karl~J.~Friston}
\author[2]{Tommaso Salvatori}
\author[3]{Takuya Isomura}
\author[2]{Alexander Tschantz}
\author[2,4]{Alex Kiefer}
\author[2]{Tim Verbelen}
\author[2]{Magnus Koudahl}
\author[2,6,9]{Aswin Paul}
\author[4]{Thomas Parr}
\author[6,7,8]{Adeel Razi}
\author[10]{Brett Kagan}
\author[2]{Christopher~L.~Buckley}
\author[1,2]{Maxwell~J.~D.~Ramstead}

\affil[1]{Wellcome Trust Centre for Neuroimaging, Institute of Neurology, University College London, UK.}
\affil[2]{VERSES AI Research Lab, Los Angeles, California, 90016, USA}
\affil[3]{Brain Intelligence Theory Unit, RIKEN Center for Brain Science, Wako, Saitama, Japan}
\affil[4]{Nuffield Department of Clinical Neurosciences, University of Oxford, UK}
\affil[5]{Monash Centre for Consciousness and Contemplative Studies, Melbourne, Australia}
\affil[6]{Turner Institute for Brain and Mental Health, School of Psychological Sciences, Monash University, Clayton, Australia}
\affil[7]{Monash Biomedical Imaging, Monash University, Clayton, Australia}
\affil[8]{CIFAR Azrieli Global Scholars Program, Toronto, Canada}
\affil[9]{IITB-Monash Research Academy, Mumbai-76, India}
\affil[10]{Cortical Labs, Melbourne, Australia}

\date{\vspace{-5ex}}

\begin{document}
\maketitle
\pagenumbering{arabic}

\begin{abstract}
    Recent advances in theoretical biology suggest that basal cognition and sentient behaviour are emergent properties of \textit{in vitro} cell cultures and neuronal networks, respectively. Such neuronal networks spontaneously learn structured behaviours in the absence of reward or reinforcement. In this paper, we characterise this kind of self-organisation through the lens of the free energy principle, i.e., as self-evidencing. We do this by first discussing the definitions of reactive and sentient behaviour in the setting of active inference, which describes the behaviour of agents that model the consequences of their actions. We then introduce a formal account of \emph{intentional} behaviour, that describes agents as driven by a preferred endpoint or goal in latent state-spaces. We then investigate these forms of (reactive, sentient, and intentional) behaviour using simulations. First, we simulate the aforementioned \textit{in vitro} experiments, in which neuronal cultures spontaneously learn to play Pong, by implementing nested, free energy minimising processes. The simulations are then used to deconstruct the ensuing predictive behaviour—leading to the distinction between merely reactive, sentient, and intentional behaviour, with the latter formalised in terms of inductive planning. This distinction is further studied using simple machine learning benchmarks (navigation in a grid world and the Tower of Hanoi problem), that show how quickly and efficiently adaptive behaviour emerges under an inductive form of active inference. 

\end{abstract}


\section{Introduction}

In 2022, a paper was published that claimed to demonstrate sentient behaviour in a neuronal culture grown in a dish (an \textit{in vitro} neuronal network) \cite{kagan2022vitro}. The behaviour in question was the spontaneous emergence of controlled movements of a paddle to hit a ball—and thereby play Pong. This study has several sources of inspiration that speak to the notion of basal cognition; see, e.g., \cite{fields2021minimal, levin2019computational, manicka2019modeling} (and related work, e.g. \cite{masumori2015emergence}). In particular, the hypothesis that adaptive and predictive behaviour would emerge spontaneously was based on earlier work showing that \textit{in vitro} neuronal cultures could be described as minimising variational free energy \cite{isomura2018vitro} and thereby evince active inference and learning. This application of the free energy principle (FEP) to neuronal cultures was subsequently validated empirically \cite{isomura2023experimental}: in the sense that changes in neuronal activity and synaptic efficacy—that underwrite learning—could be predicted quantitatively, as a variational free energy minimising process. So, are these findings remarkable, or were they predictable?

In one sense, these results were entirely predictable. Indeed, they were predictable from the FEP, which states that any two networks—that are coupled in a certain sparse fashion—will come to manifest a generalised synchrony \cite{friston2021sophisticated, palacios2019emergence}. More formally, the FEP states that if the probability density that underwrites the dynamics of coupled random dynamical systems contains a Markov blanket—which shields internal states from external states, given blanket (sensory and active) states—then internal states will look as if they track the statistics of external states—or more precisely, as if they encode the parameters of a variational density (or best guess about) external states beyond the blanket. Empirically, this synchronisation was observed when the neuronal cultures learned to play Pong. However, the FEP goes further and says that the internal and active states (together, autonomous states) of either network can be described as minimising a variational free energy functional. This functional is exactly the same used to optimise generative models in statistics and machine learning \cite{winn2005variational}. On this reading, one can interpret the autonomous states—of a network, particle or person—as minimising variational free energy or surprise (a.k.a., self-information) or, equivalently, maximising Bayesian model evidence (a.k.a., the marginal likelihood of sensory states). This leads to an implicit teleology, in the sense that one can describe self-organisation in terms of self-evidencing \cite{hohwy2016self} that entails active inference and learning, planning, purpose, intentions and, perhaps, sentience. The underlying free energy minimising processes—and their teleological interpretation—are the focus of this paper.

The results reported in \cite{kagan2022vitro} were considered by some to be unremarkable for a different reason: learning to play (Atari) games like Pong was something that had been accomplished with machine learning systems years earlier using neural networks and (deep) reinforcement learning \cite{mnih2015human, schrittwieser2019mastering}. So, what is remarkable about a neuronal network reproducing the same kind of behaviour? It is remarkable because one cannot use the reinforcement learning (RL) paradigm to explain the emergence of self-evidencing behaviour seen \textit{in vitro}. This follows from the fact that one cannot reward a neuronal network—because no one knows what any given \textit{in vitro} neuronal network finds rewarding. However, the FEP theorist knows exactly what a self-evidencing network finds aversive; namely, surprise and unpredictability. This was a rationale for delivering unpredictable noise to the sensory electrodes of the cell culture (or restarting the game in an unpredictable way), whenever the neuronal network failed to hit the ball \cite{kagan2022vitro}.

Some found the results reported in \cite{kagan2022vitro} remarkable, but not in a good way: they disagreed with the claim that the behaviour could be described as ‘sentient’ \cite{balci2023response}. Here, we hope to make sense of the notion of sentient behaviour in terms of Bayesian belief updating; where ‘sentient behaviour’ denotes the capacity to generate appropriate responses to sensory perturbations (as opposed to merely reactive behaviour). We pursue the narrative established by the cell culture experiments above to illustrate why Pong-playing behaviour was considered sentient, as opposed to reactive. In brief, we consider a bright line between actions based upon the predictions of a generative model that does, and does not, entail the consequences of action. 

Specifically, this paper differentiates between three kinds of behaviour: \emph{reactive, sentient, and intentional}. The first two have formulations that have been extensively studied in the literature, under the frameworks of model-free reinforcement learning (RL) and active inference, respectively. In model-free RL, the system selects actions using either a lookup table (Q-learning), or a neural network (deep Q-learning). In standard active inference, the action selection depends on the expected free energy of policies (Equation~\ref{eq:2}), where the expectation is over \textit{observations in the future} that become random variables. This means that preferred outcomes—that subtend expected cost and risk—are prior beliefs that constrain the implicit planning as inference \cite{attias2003planning, botvinick2012planning, van2013informational}. Things that evince this kind of behaviour can hence be described as planning their actions, based upon a generative model of the consequences of those actions \cite{attias2003planning, botvinick2012planning, dacosta2020active}. It was this sense in which the behaviour of the cell cultures was considered sentient. 

This form of sentient behaviour —described in terms of Bayesian mechanics \cite{ramstead2023bayesian, friston2023free, friston2022path}—can be augmented with intended endpoints or goals. This leads to a novel kind of sentient behaviour that not only predicts the consequences of its actions, but is also able to select them to reach a goal state that may be many steps in the future. This kind of behaviour, that we call \emph{intentional behaviour}, generally requires some form of backwards induction \cite{camerer1997progress, hure2020deep} of the kind found in dynamic programming \cite{bellman1952dynamic, dacosta2020relationship, sutton1999between, paul2023efficient}: this is, starting from the intended goal state, and working backwards, inductively, to the current state of affairs, in order to plan moves to that goal state. Backwards induction was applied to the partially observable setting and explored in the context of active inference in \cite{paul2023efficient}. In that work, dynamic programming was shown to be more efficient than traditional planning methods in active inference.

The focus of this work is to formally define a framework for intentional behaviour, where the agent minimises a constrained form of expected free energy—and to demonstrate this framework \textit{in silico}. These constraints are defined on a subset of latent states that represent the intended goals of the agent, and propagated to the agent via a form of backward induction. As a result, states that do not allow the agent to make any `progress' towards one of the intended goals are penalised, and so are actions that lead to such disfavoured states. This leads to a distinction between sentient and intentional behaviour, were intentional behaviour is equipped with inductive constraints.

In this treatment, the word \emph{inductive} is used in several senses. First, to distinguish inductive planning from the abductive kind of inference that usually figures in applications of Bayesian mechanics; i.e., to distinguish between mere inference to the best explanation (abductive inference) and genuinely goal-directed inference (inductive planning) \cite{harman1965inference, seth2015inference}. Second, it is used with a nod to backwards induction in dynamic programming, where one starts from an intended endpoint and works backwards in time to the present, to decide what to do next \cite{bellman1952dynamic, dacosta2020relationship, howard1960dynamic, paul2023efficient}. Under this naturalisation of behaviours, a thermostat would not exhibit sentient behaviour, but insects might (i.e., thermostats exhibit merely reactive behaviour). Similarly, insects would not exhibit intentional behaviour, but mammals might (i.e., insects exhibit merely sentient behaviour). The numerical analyses presented below suggest that\textit{ in vitro} neuronal cultures may exhibit sentient behaviour, but not intentional behaviour. Crucially, we show that neither sentient nor intentional behaviour can be explained by reinforcement learning. In the experimental sections of this work, we study and compare the performance of active inference agents with and without intended goal states. For ease of reference, we will call active inference agents without goal states \emph{abductive agents}, and agents with intended goals \emph{inductive agents}.

This paper comprises four sections. The first briefly rehearses active inference and learning—as a set of nested free energy minimising processes—applied to a generic generative model of exchange with some world or environment. This model is a partially observed Markov decision process that is conciliatory with canonical neural networks in machine learning and apt to describe the self-evidencing of in vitro neuronal networks \cite{isomura2018vitro, isomura2023experimental}. This section has a special focus on inductive planning and its relationship to expected free energy. The subsequent sections use numerical studies to make a series of key points. The second section reproduces the empirical behaviour of \textit{in vitro} neuronal networks playing Pong. Crucially, this behaviour emerges purely in terms of free energy minimising processes, starting with a naïve neuronal network. This section illustrates the failure of a (simulated) abductive agent when the game is made more difficult. This failure is used to illustrate the role of inductive planning, which restores performance and underwrites a fluent engagement with the sensorium. The final two sections illustrate inductive planning using navigation in a maze and the Tower of Hanoi problem, respectively. These numerical studies illustrate how the simple application of inductive constraints to active inference allows tasks—that would be otherwise intractable in discrete state spaces—to be solved efficiently. This efficiency rests on the fact that distal goals can be reached by only planning  a few steps in the future, thanks to constraints furnished by inductive planning. Effectively, inductive planning takes the pressure off deep tree searches by identifying 'blind alleys' or 'dead ends'.

\subsection{Glossary of definitions}

Before introducing the inductive planning algorithm, we frame our treatment by clarifying our use of some key terms. This framing is important, given that the goal of the present work is not simply to describe a useful heuristic for efficient inference (i.e., inductive planning), but to provide an account of how a new form of decision-making, characteristic of more complex forms of agency, may be combined with, and folded into, a generic Bayesian (active) inference scheme.

\begin{figure}[t!]
    \centering
    \includegraphics[width=1.0\textwidth]{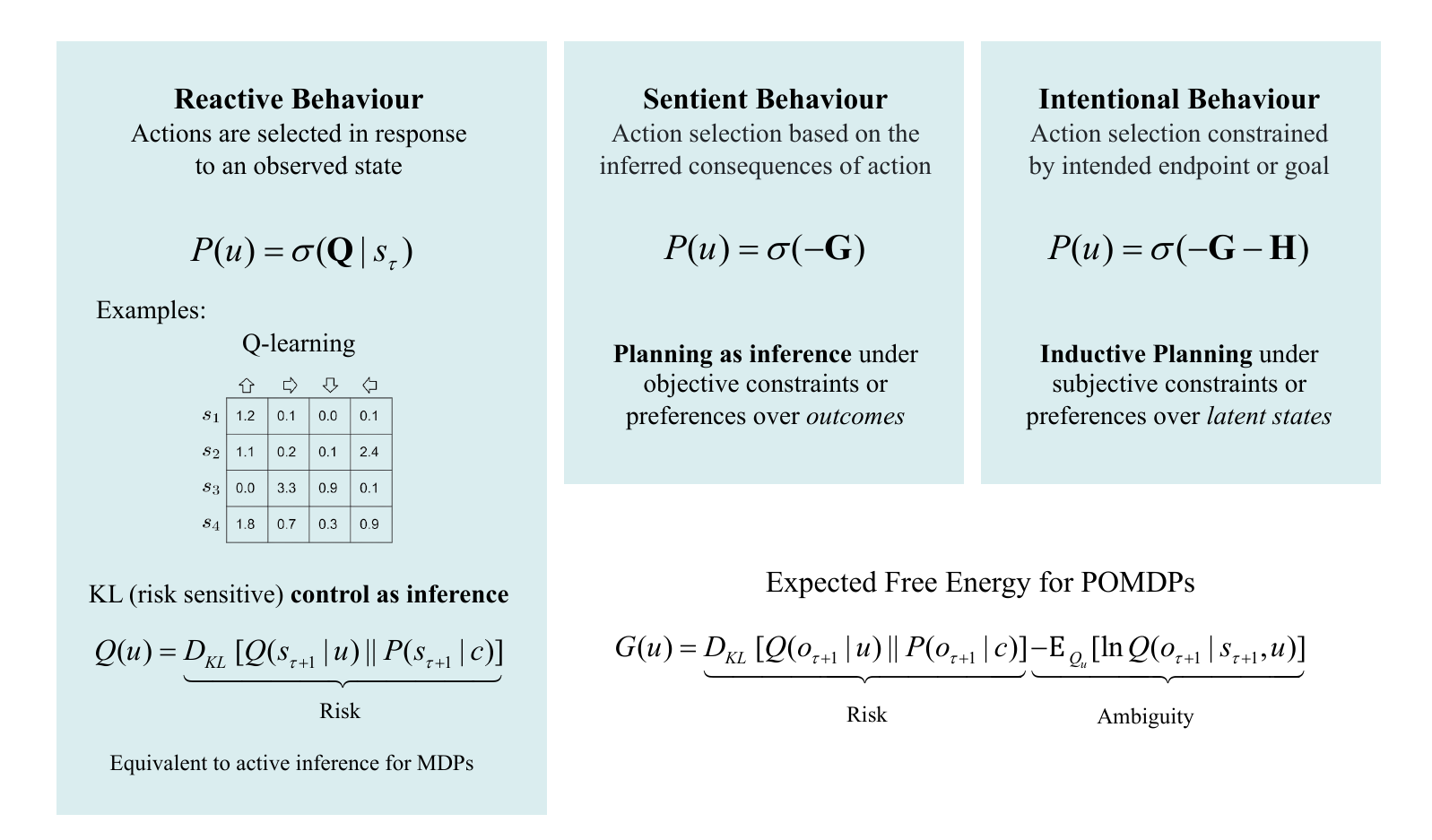}
    \caption{\textbf{Glossary.} In this figure, we provide illustrative definitions of the three kinds of behaviour considered in this work, In terms of examples, and mathematical differences. Examples of agents with reactive behaviours are (1) Model-free reinforcement learning schemes, such as Q-learning, where the agent makes use of a lookup table to select actions (more generally, a state-action policy). In this table, rows correspond to states, actions to columns, and every entry encodes the value of taking a specific action  (in this case: go up, right, down, left) when in state $s_{\tau}$. There is no inference over policies, as for every state the agent automatically selects the action with the highest value; and (2) KL control (a.k.a., risk-sensitive control) methods, that automatically select actions that minimise a KL divergence between anticipated and preferred states (where there is no uncertainty about the current state). Sentient agents, on the other hand, plan by taking into account future outcomes and their uncertainty, as they act by minimising an expected free energy $\mathbf{G}$, that includes risk and ambiguity terms. More details on this can be found in Equation~\ref{eq:3}. Finally, inductive agents add constraints ($\mathbf{H}$ in the figure) in the action selection, by penalising actions that preclude an intended goal. For a formal derivation of $\mathbf{H}$, we refer to Section~\ref{sec:ind}.}
\label{fig:glossary}
\end{figure}

Figure~\ref{fig:glossary} describes increasingly complex forms of behaviour---from reactive (merely responding to stimuli), to sentient (planning based on the \textit{sensory consequences} of actions), to intentional (planning in order to bring about \textit{intended states})---and corresponding forms of decision-making that may underwrite such behaviour. 

\textbf{Reactive behaviour} characterises simple sensorimotor reflex arcs and the mere realisation of set points or trajectories (e.g., simple cases of homeostasis and homeorhesis). This form of behaviour can be accounted for acting in a way that realises predicted sensations, with no anticipation of the future sensory consequences of action.

\textbf{Sentient behaviour} characterises the paradigmatic case of active inference, in which the influence of perception on action is mediated by the results of planning, with a distribution over policies derived from a model endowed with counterfactual depth (i.e., beliefs about the future sensory consequences of action pursuant to a policy). In this case, we may characterise the form of inference over actions or policies as \textit{abductive}---i.e., as an inference to the policy that best explains current and future observations under a generative model (see below).

\textbf{Intentional behaviour} is driven not simply by the generic imperative to minimise sensory prediction error, present and future, but toward the attainment of a particular future endpoint or goal state. This form of behaviour can be subserved by backward induction or \textit{inductive planning}, as defined below, which supplies a specific form of constraint on the Bayesian (abductive) inference characteristic of (mere) sentient behaviour. In particular, it implies not merely beliefs about sensory consequences of actions but rather beliefs about the inferred or latent causes of sensory input. 

Note that words like ‘sentient behaviour’ and ‘intentional behaviour’ are deliberately defined here such that they can be operationalized within the framework of generative modelling, in which terms like ‘state’, ‘belief’, and ‘confidence’ have precise, if narrow, interpretations in terms of belief structures of a mathematical sort \cite{ramstead2022bayesian}. Whether the phenomenology of (propositional or subjective) beliefs—or sentience—could yield to the same naturalisation remains to be seen: see \cite{clark2019bayesing,sandved-smith2021towards,Smith2022-SMIAIM-4} for treatments in this direction. Note further that a key distinction between sentient and intentional behaviour rests upon the consequences of behaviour in (observable) outcome and (unobservable) latent spaces, respectively.

\section{Active inference}\label{sec:active_inference}

Here, we introduce the generative model used in the following sections, which can be seen as a generalisation of a partially observed Markov decision process (POMDP). The generalisation in question covers trajectories, narratives or syntax—which may or may not be controllable—by equipping a POMDP with random variables called \textit{paths}. Paths effectively pick out transitions among latent states. These models are designed to be composed hierarchically, in a way that speaks to a separation of temporal scales in deep generative models. In other words, the number of transitions among latent states at any given level is greater than the number of transitions at the level above. This furnishes a unique specification of a hierarchy, in which the parents of any latent \textit{factor} (associated with unique states and paths) contextualise the dynamics of their children.

The variational inference scheme \cite{beal2003variational} used to invert these models inherits from their application to online decision-making tasks. This means that action selection rests primarily on current beliefs about latent states and structures, and expectations about future observations. In that sense, the beliefs are updated sequentially—and in an online fashion—with each new action-outcome pair. This calls for Bayesian filtering (i.e., forward message passing) during the active sampling of observations, followed by Bayesian smoothing (i.e., forward and backward message passing) to revise posterior beliefs about past states at the end of an epoch. The implicit Bayesian smoothing ensures that the beliefs about latent states at any moment in the past are informed by all available observations when updating model parameters (and latent states of parents in deep models). 

In neurobiology, this combination of Bayesian filtering and smoothing would correspond to evidence accumulation during active engagement with the environment, followed by a ‘replay’ before the next epoch \cite{buckner2010role, louie2001temporally, penny2013forward, pezzulo2014internally}. From a machine learning perspective, this can be regarded as a forward pass (c.f., belief propagation) for online active inference, followed by a backwards pass (implemented with variational message passing) for active learning. The implicit belief updates, pertaining to states, parameters and structure, foreground the conditional dependencies between active inference, learning, and selection, respectively.

\subsection*{Generative modelling}

Active inference rests upon a \textit{generative model} of observable outcomes (observations). This model is used to infer the most likely causes of outcomes in terms of expected states of the world. These states (and paths) are latent or \textit{hidden} because they can only be inferred through observations. Some paths are controllable in the sense they can be realised through action. Therefore, certain observations depend upon action (e.g., where one is looking), which requires the generative model to entertain expectations about outcomes under different combinations of actions (i.e., policies)\footnote{Note that in this setting, a policy is not a sequence of actions, but simply a combination of paths, where each hidden factor has an associated state and path. This means there are, potentially, as many policies as there are combinations of paths.}. 

These expectations are optimised by minimising the \textit{variational free energy}, defined in Equation~\eqref{eq:1}. Variational free energy scores the discrepancy between the data expected under the generative model and the actual data. Crucially, the prior probability of a policy depends upon its \textit{expected free energy}. Expected free energy, described in more detail in Equation~\eqref{eq:2}, is a universal objective function that can be read as augmenting mutual information with a expected costs or constraints that need to be satisfied. Heuristically, it scores the free energy expected under each course of action. Having evaluated the expected free energy of each policy, the most likely action can be selected and the perception-action cycle continues \cite{parr2022active}.

\subsection*{The generative model}

\begin{figure}[t!]
    \centering
    \includegraphics[width=0.6\textwidth]{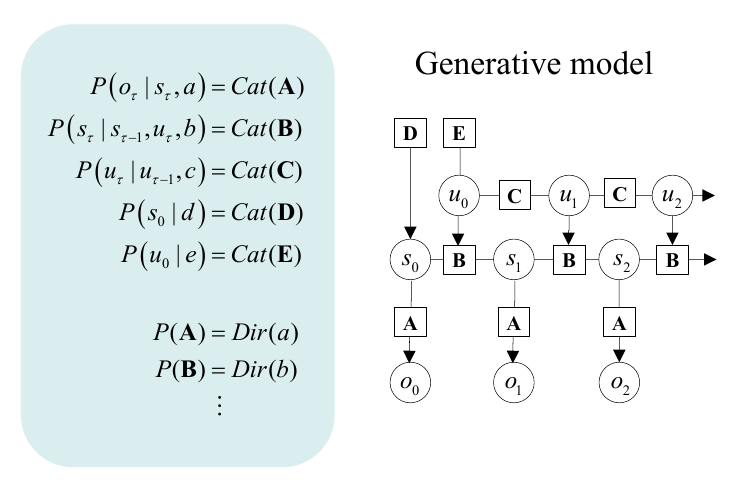}
    \caption{\textbf{Generative models as agents.} A generative model specifies the joint probability of observable consequences and their hidden causes. Usually, the model is expressed in terms of a \textit{likelihood} (the probability of consequences given their causes) and \textit{priors} (over causes). When a prior depends upon a random variable it is called an \textit{empirical prior}. Here, the likelihood is specified by a tensor $\mathbf{A}$, encoding the probability of an outcome under every combination of \textit{states} ($s$). The empirical priors pertain to transitions among hidden states, $\mathbf{B}$, that depend upon \textit{paths} ($u$), whose transition probabilities are encoded in $\mathbf{C}$. $\mathbf{E}$ specifies the empirical prior probability of each path. The subscripts in this graphic pertain to time.}
\label{fig:aif_agent}
\end{figure}

Figure \ref{fig:aif_agent} provides a schematic overview of the generative model used for the simulations considered in this paper. Outcomes at any particular time depend upon hidden \textit{states}, while transitions among hidden states depend upon \textit{paths}. Note that paths are random variables, in the sense that a particle can have both a position (i.e., a state) and momentum (i.e., a path). Paths may or may not depend upon action. The resulting POMDP is specified by a set of tensors. The first set of parameters, denoted $\mathbf{A}$, maps from hidden states to outcome modalities; for example, exteroceptive (e.g., visual) or proprioceptive (e.g., eye position) \textit{modalities}. These parameters encode the likelihood of an outcome given their hidden causes. The second set $\mathbf{B}$ prescribes transitions among the hidden states of a \textit{factor}, under a particular path. Factors correspond to different kinds of causes; e.g., the location versus the class of an object. The remaining tensors encode prior beliefs about paths $\mathbf{C}$, and initial states $\mathbf{D}$. The tensors—encoding probabilistic mappings or contingencies—are generally parameterised as Dirichlet distributions, whose sufficient statistics are concentration parameters or \textit{Dirichlet counts}. These count the number of times a particular combination of states or outcomes has been inferred. We will focus on learning the likelihood model, encoded by Dirichlet counts, $\boldsymbol{a}$.

The generative model in Figure \ref{fig:aif_agent} means that outcomes are generated as follows: first, a policy is selected using a softmax function of expected free energy. Sequences of hidden states are generated using the probability transitions specified by the selected combination of paths (i.e., policy). Finally, these hidden states generate outcomes in one or more modalities. Perception or inference about hidden states (i.e., state estimation) corresponds to inverting a generative model, given a sequence of outcomes, while learning corresponds to updating model parameters. Perception therefore corresponds to updating beliefs about hidden states and paths, while learning corresponds to accumulating knowledge in the form of Dirichlet counts. The requisite expectations constitute the sufficient statistics $(\mathbf{s}, \mathbf{u}, \mathbf{a})$ of posterior beliefs $Q(s, u, a) = Q_{\mathbf{s}}(s)Q_{\mathbf{u}}(u)Q_{\mathbf{a}}(a)$. The implicit factorisation of this approximate posterior effectively partitions model inversion into inference, planning, and learning.

\subsection*{Variational free energy and inference}

In variational Bayesian inference (a form of approximate Bayesian inference), model inversion entails the minimisation of variational free energy with respect to the sufficient statistics of approximate posterior beliefs. This can be expressed as follows, where, for clarity, we will deal with a single factor, such that the policy (i.e., combination of paths) becomes the path, $\pi = u$. Omitting dependencies on previous states, we have for model $m$:

\begin{equation}
\begin{aligned}
Q\left(s_\tau, u_\tau, a\right) & =\arg \min _Q F \\
F & =\mathbb{E}_Q[\ln \underbrace{Q\left(s_\tau, u_\tau, a\right)}_{\text {posterior }}-\ln \underbrace{P\left(o_\tau \mid s_\tau, u_\tau, a\right)}_{\text {likelihood }}-\ln \underbrace{P \left(s_\tau, u_\tau, a\right)}_{\text {prior }}] \\
& =\underbrace{D_{K L}\left[Q\left(s_\tau, u_\tau, a\right) \| P\left(s_\tau, u_\tau, a \mid o_\tau\right)\right]}_{\text {divergence }}-\underbrace{\ln P\left(o_\tau \mid m\right)}_{\text {log evidence }} \\
& =\underbrace{D_{K L}\left[Q\left(s_\tau, u_\tau, a\right) \| P\left(s_\tau, u_\tau, a\right)\right]}_{\text {complexity }}-\underbrace{\mathbb{E}_Q\left[\ln P\left(o_\tau \mid s_\tau, u_\tau, a\right)\right]}_{\text {accuracy }}
\label{eq:1}
\end{aligned}
\end{equation}

Because the (KL) divergences cannot be less than zero, the penultimate equality means that free energy is minimised when the (approximate) posterior is equal to the true posterior. At this point, the free energy is equal to the negative log evidence for the generative model \cite{beal2003variational}. This means minimising free energy is mathematically equivalent to maximising model evidence, which is, in turn, equivalent to minimising the complexity of accurate explanations for observed outcomes.

Planning emerges under active inference by placing priors over (controllable) paths to minimise expected free energy \cite{friston2015active}:
\begin{align}
\label{eq:2}
G(u) &=\mathbb{E}_{Q_{u}}\left[\ln Q\left(s_{\tau+1}, a \mid u\right)-\ln Q\left(s_{\tau+1}, a \mid o_{\tau+1}, u\right)-\ln P\left(o_{\tau+1} \mid c\right)\right] \nonumber \\ \\
& =-\underbrace{\mathbb{E}_{Q_{u}}\left[\ln Q\left(a \mid s_{\tau+1}, o_{\tau+1}, u\right)-\ln Q(a \mid s_{\tau+1}, u)\right]}_{\text {expected information gain (learning) }}- \nonumber \\
& \ \ \ \ \ \underbrace{\mathbb{E}_{Q_{u}}\left[\ln Q\left(s_{\tau+1} \mid o_{\tau+1}, u\right)-\ln Q\left(s_{\tau+1} \mid u\right)\right]}_{\text {expected information gain (inference) }} \underbrace{-\mathbb{E}_{Q_{u}}\left[\ln P\left(o_{\tau+1} \mid c\right)\right]}_{\text {expected cost }}  \\ \\
& =-\underbrace{\mathbb{E}_{Q_{u}}\left[D_{K L}\left[Q\left(a \mid s_{\tau+1}, o_{\tau+1}, u\right) \| Q(a \mid s_{\tau+1}, u)\right]\right]}_{\text {novelty }}+ \nonumber \\
& \ \ \ \ \ \underbrace{D_{K L}\left[Q\left(o_{\tau+1} \mid u\right) \| P\left(o_{\tau+1} \mid c\right)\right]}_{\text {risk }} \underbrace{-\mathbb{E}_{Q_{u}}\left[\ln Q\left(o_{\tau+1} \mid s_{\tau+1}, u\right)\right]}_{\text {ambiguity }} \nonumber
\end{align}

Here, the posterior predictive distribution over parameters, hidden states and outcomes at the next time step, under a particular path, is defined as follows:
\begin{align*}
 Q_u &= Q\left(o_{\tau+1}, s_{\tau+1}, a \mid u\right) \\ &= P\left(o_{\tau+1}, s_{\tau+1}, a \mid u, o_0, \ldots, o_\tau\right) \\ &= P\left(o_{\tau+1} \mid s_{\tau+1}, a\right) Q\left(s_{\tau+1}, a \mid u\right).   
\end{align*}

One can also express the prior over the parameters in terms of an expected free energy, where, marginalising over paths:

\begin{equation}
\begin{aligned}
P(a) & =\sigma(-G) \\\\
G(a) & =\mathbb{E}_{Q_a}[\ln P(s \mid a)-\ln P(s \mid o, a)-\ln P(o \mid c)] \\\\
& =-\underbrace{\mathbb{E}_{Q_a}[\ln P(s \mid o, a)-\ln P(s \mid a)]}_{\text {expected information gain }} \underbrace{-\mathbb{E}_{Q_a}[\ln P(o \mid c)]}_{\text {expected cost }} \\ 
& =-\underbrace{\mathbb{E}_{Q_a}\left[D_{K L}[P(o, s \mid a) \| P(o \mid a) P(s \mid a)]\right.}_{\text {mutual information }} \underbrace{-\mathbb{E}_{Q_a}[\ln P(o \mid c)]}_{\text {expected cost }}
\label{eq:3}
\end{aligned}
\end{equation}

where $Q_a = P(o|s, a) P(s|a) = P(o, s|a)$ is the joint distribution over outcomes and hidden states, encoded by the Dirichlet parameters, $a$, and $\sigma(\cdot)$ is the softmax function. Note that the Dirichlet parameters encode the mutual information, in the sense that they implicitly encode the joint distribution over outcomes and their hidden causes. When normalising each column of the $a$ tensor, we recover the likelihood distribution (as in Figure \ref{fig:aif_agent}); however, we could normalise over every element, to recover a joint distribution.

As discussed above, expected free energy can be regarded as a universal objective function that augments mutual information with expected costs or constraints. Constraints — parameterised by $c$ — reflect the fact that we are dealing with open systems with characteristic outcomes. This allows an optimal trade-off between exploration and exploitation, that can be read as an expression of the constrained maximum entropy principle that is dual to the free energy principle \cite{ramstead2023bayesian}. Alternatively, it can be read as a constrained principle of maximum mutual information or minimum redundancy \cite{ay2008predictive, barlow1961possible, linsker1990perceptual, olshausen1996emergence}. In machine learning, this kind of objective function underwrites disentanglement \cite{higgins2021unsupervised, sanchez2020learning}, and generally leads to sparse representations \cite{gros2009cognitive, olshausen1996emergence, sakthivadivel2022weak, tipping2001sparse}.

When comparing the expressions for expected free energy in Equation~\ref{eq:2} with variational free energy in Equation~\ref{eq:1}, the expected divergence becomes expected information gain. Expected information gain about the parameters and states are sometimes associated with distinct epistemic affordances; namely, \textit{novelty} and \textit{salience}, respectively \cite{schwartenbeck2019computational}. Similarly, expected log evidence becomes expected value, where value is the logarithm of prior preferences. The last equality in Equation~\ref{eq:2} provides a complementary interpretation; in which the expected complexity becomes risk, while expected inaccuracy becomes ambiguity.

There are many special cases of minimising expected free energy. For example, maximising expected information gain maximises (expected) Bayesian surprise \cite{itti2009bayesian}, in accord with the principles of optimal (Bayesian) experimental design \cite{lindley1956measure}. This resolution of uncertainty is related to artificial curiosity \cite{schmidhuber1991curious, still2012information} and speaks to the value of information \cite{howard1966information}.

Expected complexity or risk is the same quantity minimised in risk sensitive or KL control \cite{klyubin2005empowerment, broek2012risk}, and underpins (free energy) formulations of bounded rationality based on complexity costs \cite{braun2011path, ortega2013thermodynamics} and related schemes in machine learning; e.g., Bayesian reinforcement learning \cite{ghavamzadeh2015bayesian}. More generally, minimising expected cost subsumes Bayesian decision theory \cite{berger2013statistical}.

\newpage

\begin{figure}[H]
    \centering
    \includegraphics[width=\textwidth]{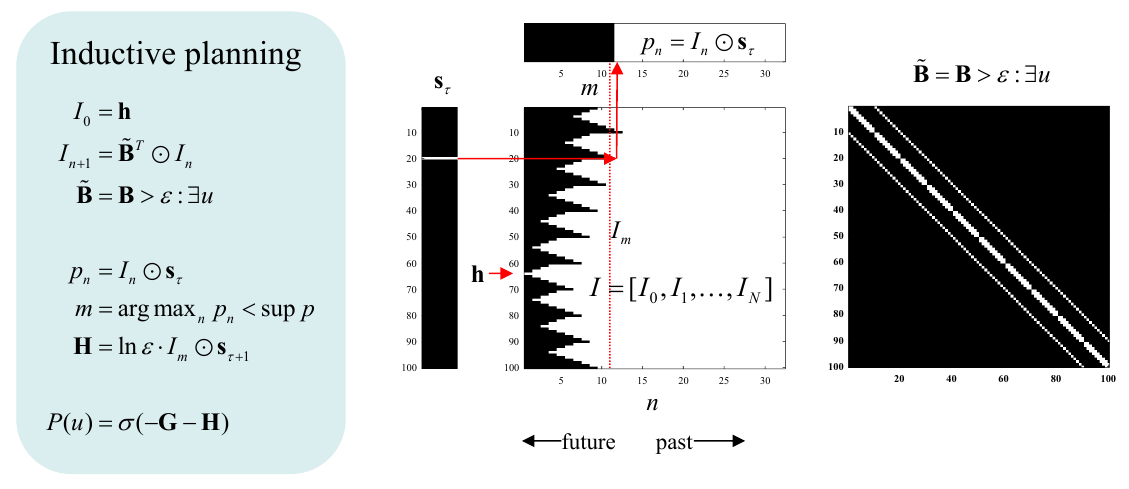}
    \caption{\textbf{Inductive Planning.} This figure provides an overview of inductive planning used in this paper. The left panel provides the expressions used to induce which subsequent states do and do not contain paths to some intended end state, encoded by a one hot vector $\mathbf{h}$. The central panel illustrates this induction graphically, where vectors and matrices are shown in image format (black equals zero or false and white equals one or true). 
    Working  down the equalities in the left panel, we first initialise a logical vector of states, $I$, to the intended state $\mathbf{h}$. Recursively, we evaluate all the states from which the previous state can be accessed (a state can be accessed if the probability of transitioning from an adjacent state is larger than $\varepsilon$). Because this recursive induction works backwards in time, the allowable transition matrix is transposed. Having induced the reverse history of states—that contain paths to the intended state—one can then evaluate the length of the shortest path to the intended state. This depends upon posterior beliefs about the current state. In the example shown on the left, we are currently in state $20$, which means that the shortest path to the intended state (state $64$) is $12$ time steps. This tells us that if we are pursuing the shortest path then there are certain states we need to avoid—from which the intended state cannot be reached. These states are encoded by the logical vector I at the next time step; namely, the last time before the probability $p$ of being on a path to the intended state reaches its supremum. Because the eligible states can only increase—as we move backwards in time—this probability can only increase, until all states are eligible (or there are no further eligible states). The first time that the probability reaches its supremum tells us where we are on the path to our intended state and, crucially, the ineligible states at the next time step. We now know the states to avoid at the next time step. If ineligible states are precluded, the next state must be on the path to the intended state. Ineligible states can be assigned a high cost (here, the log of a small value) to evaluate the expected cost incurred by each policy, using its predictive posterior over states (see Figure~\ref{fig:aif_agent}). Finally, we can supplement the expected free energy, $\mathbf{G}$ of each policy with the ensuing inductive cost, $\mathbf{H}$. In principle, this guarantees the selection of paths or policies that lead to the intended state, provided that state can be reached. The example shown on the right is taken from the maze navigation task described later. For clarity, this example only considers a single factor. The mathematical expressions use the notation of Figure~\ref{fig:aif_agent}: The dotted red line indicates the logical vector encoding which of the $100$ states will lead to the intended state at the next time point; here, $11$ time steps from the intended state (indicated with a small red arrow). }
\label{fig:inductive}
\end{figure}

\newpage

\section{Inductive Planning}\label{sec:ind}

What we call inductive planning—in this setting—recalls the notion of backwards induction in dynamic programming and related schemes \cite{camerer2004cognitive, dacosta2020relationship, howard1960dynamic, hure2020deep, sutton1999between, tervo2016toward, paul2023efficient}.  In this form of inference, precise beliefs about state transitions are leveraged to rule out actions that are inconsistent with the attainment of future goals, defined in belief or state space as a final (or intended) state. This is a limiting case of inductive (Bayesian) inference \cite{barlow1974inductive, hawthorne2021inductive, Kiefer2017-KIELPI} in which the very high precision of beliefs about final or intended states allows one to use logical operators in place of tensor operations; thereby vastly simplifying computations. In brief, we will use this simplification to furnish constraints on action selection that inherit from priors over intended states in the future. 

Active inference rests on priors that place constraints on paths or trajectories through state space. For example, a sparse prior preference with knowledge only about the final state warrants deep planning to demonstrate intentional behaviour \cite{paul2023efficient}. One can either specify these constraints in terms of states that are unlikely to traversed, or in terms of the final state. In other words, the agent may, \textit{a priori}, believe it will navigate state space in a way that avoids unlikely or surprising outcomes, or that it will reach some final destination (in state space, not outcome space), irrespective of the path taken. These are distinct kinds of constraints. The first is implemented by $\mathbf{c}$, in terms of the cost or constraints that apply during the entire path. We now introduce another prior or constraint $\mathbf{h}$, over the final state. The priors, $\mathbf{d}$ and $\mathbf{h}$ play reciprocal roles; in the sense they specify prior beliefs about the initial and final states, respectively. Backwards induction now follows simply from this prior; provided it is specified sufficiently precisely. We will refer to these final states as intended states\footnote{While $\mathbf{c}$, $\mathbf{d}$, and $\mathbf{h}$ are usually hard coded, they can be learnt very efficiently, for example using Z-learning for certain classes of MDPs \cite{todorov2006linearly,paul2023efficient}}.

The basic idea is that although we may be uncertain about the next latent state, we can be certain about which states cannot be accessed from the current state. This means we can use induction to identify subsequent states that cannot be on a path to an intended state; thereby rendering actions—(i.e., state transitions) to those ineligible, 'dead-end' states—highly unlikely (assuming that we are on a path to an intended state). The requisite induction goes as follows:

Imagine that we know our current state and that we will be in a certain (intended) state in the future. Imagine further that we know all possible transitions, afforded by action, among states. This means we can identify all the states from which the intended state is accessible. We can now repeat this and identify all the states from which the eligible states at the penultimate time point can be accessed, and so on. We now repeat this recursively—moving backwards in time—until our current state becomes eligible. At this point, we select an action that precludes ineligible states at the preceding point in backwards time (or next point in forwards time), bringing us one step closer to the intended state. We now repeat the backwards induction, until we arrive at the intended state, via the shortest path. This backwards induction is computationally cheap because it entails logical operations on a sparse logical tensor, encoding allowable state transitions.

Figure~\ref{fig:inductive} provides a pseudocode and graphical abstraction based upon the MATLAB scripts implementing this inductive logic. For clarity, we have assumed a single factor and that there are no constraints on the paths, other than those specified by a one hot vector $\mathbf{h}$, specifying the agent’s intended states\footnote{In our MATLAB implementation of inductive planning, constraints due to prior preferences in outcome space are accommodated by precluding transitions to costly states during construction of the logical matrix encoding possible or true transitions. Furthermore, the implementation deals with multiple factors using appropriate tensor products. Finally, when multiple intended states are supplied, the nearest state is chosen for induction; where nearest is defined in terms of the number of timesteps required to access an intended state.}.

Note that this is not vanilla backwards induction. It is simply a way of placing precise priors on paths that render certain paths—that cannot access an intended state—highly unlikely. The requisite priors complement expected free energy in the following sense (see Figure 3): inductive priors over policies, $\mathbf{H}$ are derived from priors over intended states $\mathbf{h}$, while the priors over policies scored by expected free energy, $\mathbf{G}$ inherit from priors over preferred outcomes $\mathbf{c}$. This distinction is important because it means that this kind of reasoning—and intentional behaviour—can only manifest under precise beliefs about latent states. For example, a baby (or unexplainable neural network) could not, by definition, act intentionally because it does not have a precise generative model of latent states (or any mechanism to specify intended states). We will return to prerequisites for inductive planning in the discussion.

In summary, inductive planning propagates constraints backwards in time to provide empirical priors for planning as inference in the usual way. This means that—within the constraints afforded by such planning—actions will still be chosen that maximise expected information gain and any constraints encoded by $\mathbf{c}$. In this sense, the inductive part of this inference scheme can be regarded as providing a constrained expected free energy, which winnows trajectories through state space to paths of least action. An equivalent and alternative perspective is that inductive planning furnishes an empirical prior over policies. 

When intended states are conditioned on some context—inferred by a supraordinate (hierarchical) level—one has the opportunity to learn intended states and, effectively, make planning habitual. In this setting, the implicit Dirichlet counts in $\mathbf{h}$, could be regarded as accumulating habitual courses of action that are learned as empirical priors in hierarchical models. We will pursue this elsewhere. In what follows, we focus on the distinction between sentient behaviour—based upon expected free energy—and intentional behaviour—based upon inductive priors.

\section{Pong Revisited}

In this section, we first simulate 'mere' sentient Behaviour and then examine the qualitative differences in behaviour when adding inductive constraints. Specifically, we simulate the \textit{in vitro} experiments reported in \cite{kagan2022vitro}, using both an abductive and an inductive agent. The first has no intended goals, and stands in for a naïve neuronal culture; the second has as set of intended states: the ones where the paddle hits upcoming balls. As environments, we use Pong of two different sizes, that reflect two different difficulties: $5 \times 6$ (easy), and $8 \times 4$ (hard). The results show that while the simulated \textit{in vitro} agent is able to fluently play in the easy environment, it struggles in the harder one. The inductive agent, on the other hand, can master the harder environment in less than three minutes of (simulated) game time. 

\begin{figure}[H]
    \centering
    \includegraphics[width=0.95\textwidth]{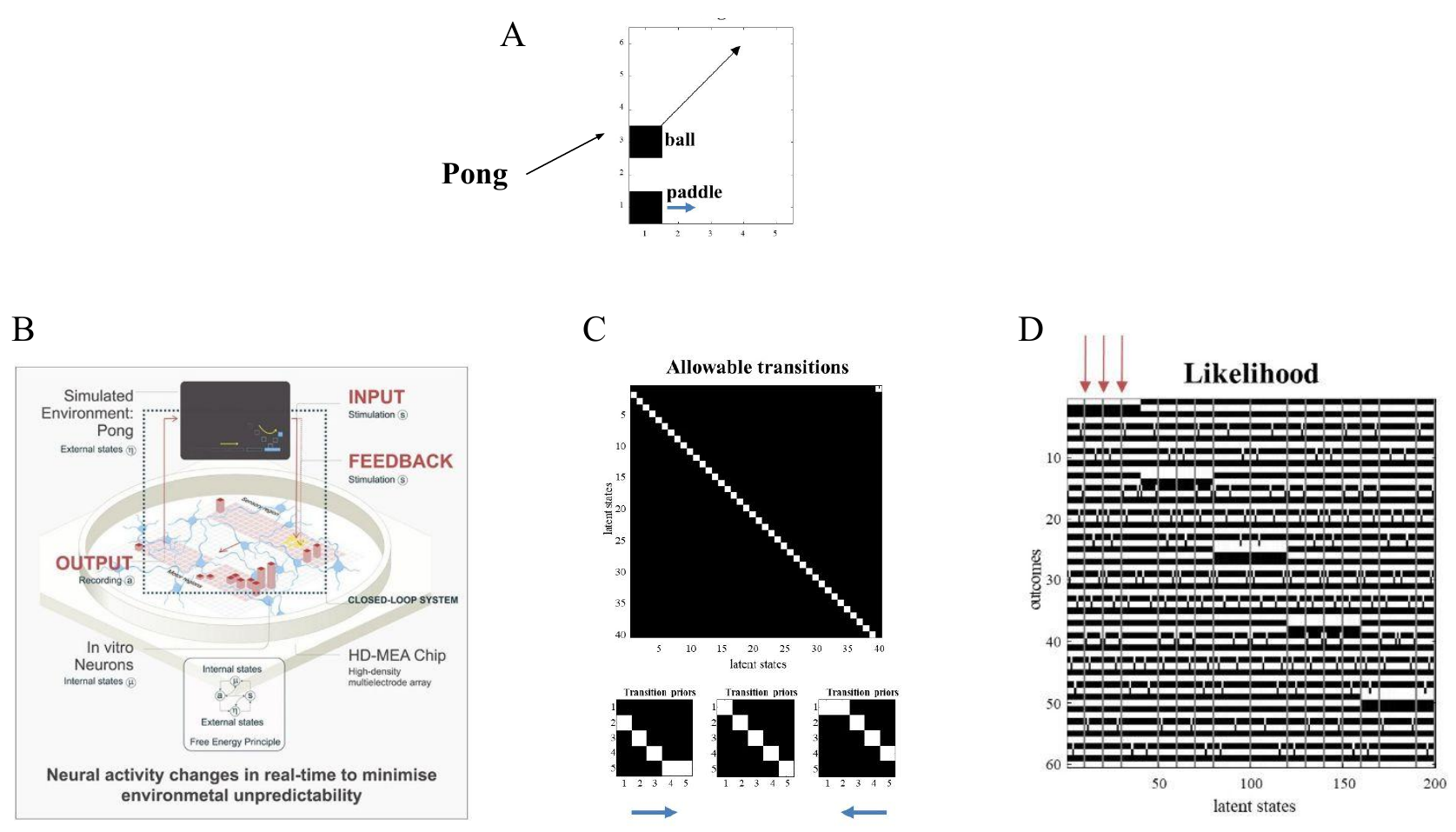}
    \caption{\textbf{Learning the world of Pong.} Panel $A$: Setup used in the simulations. In brief, the generative process modelled a ball bouncing around inside a bounding box, with a movable paddle on the lower boundary. The $(5 \times 6 =)  30$ locations or pixels provided outputs with two states (black or white) that were subsequently learned via a likelihood mapping to $40$ latent states. The agent was equipped with a precise transition prior where $40$ latent states followed each other, with circular boundary conditions. In addition, the agent was equipped with a second factor that controlled the panel, moving it to the right, staying still and moving it to the left. Panel $B$: graphical abstract (reproduced with permission from the authors) describing the \textit{in vitro} empirical study in which a closed loop system was used to record from—and stimulate—a network of cultured neurons. The set up enabled the neurons to control a virtual paddle in a simulated game of Pong. Sensory feedback reported the location of the ball and paddle; enabling the neuronal preparation to learn how to play a rudimentary form of ping-pong. Panel $C$ shows the transitions of the generative model, while Panel $D$ shows the results of active learning—i.e., accumulation of Dirichlet counts in the likelihood tensor—after $512$ time steps. Note that this is a precise likelihood mapping due to the fact that the synthetic agent has precise, if generic, transition priors. The likelihood mapping in panel D is shown in image format, with each of the $30$ likelihood tensors stacked on top of each other. Of note here are certain latent states that produce ambiguous (i.e., unpredictable) outcomes. The first three are labelled with small arrows over the likelihood matrix. These ambiguous likelihood mappings appear as grey columns. This reflects the fact that the agent has learned that states corresponding to ‘missing the ball’ lead to unpredictable and ambiguous stimulation. The implicit surprise and ambiguity means that the agent plans to avoid these states and look as if it is playing Pong—by choosing paths or policies that are more likely to hit the ball. The emergence of this behaviour is described in the next figure.}
\label{fig:pong}
\end{figure}

In the \textit{in vitro} experiments, certain cells were stimulated depending upon the configuration of a virtual game of Pong, constituted by the position of a paddle and a ball bouncing around a bounded box. Other recording electrodes were used to drive the paddle, thereby closing the sparse coupling between the neuronal network and the computer network simulating the game of Pong (see Figure~\ref{fig:pong}). Typically, in these experiments, after a few minutes of exposure to the game, short rallies of ball returns emerge. To emulate this setup, we created a generative process (i.e., a hard-coded representation of the dynamics of external states) in which a ball bounced around a box at $45$ degrees. The lower boundary contained a paddle that could be moved to the right or left. The size of the box was $5 \times 6$ units, where the ball moved one unit up or down (and right or left) at every time point. The (one unit wide) paddle could be moved left or right by one unit at every time point. In the in vitro experiments, whenever the agent missed the ball, either white noise or no stimulation was applied to the sensory electrodes; otherwise, the game remained in the play. We simulated this by supplying random input to all sensory channels whenever the ball failed to contact the paddle on the lower boundary.

The (sensory) outcomes of the POMDP comprised $30$ sensory channels that could be on or off. These can be thought of as pixels in a simple Atari-like game. The latent states were modelled as one long orbit, by equipping the generative model with a transition matrix that moved from one state to the next (with circular boundary conditions) for a suitably long sequence of state transitions (here, $40$). The generative model was equipped with a second factor with three controllable paths. This factor moved the paddle one unit to the right or left (or no movement). However, the (implicit) agent knew nothing more about its world and, in particular, had no notion that the second factor endowed it with control over the paddle. This was because the likelihood tensors mapping from the two latent factors to the outcomes were populated with small and uniform Dirichlet counts (i.e., concentration parameters of $1/32$). In other words, our naïve generative model could, in principle, model any given world (providing this world has a limited number of states that are revisited systematically). Figure~\ref{fig:pong} shows the setup of this paradigm and the parameters of the generative model learned after $512$ time steps.

\begin{figure}[H]
    \centering
    \includegraphics[width=0.85\textwidth]{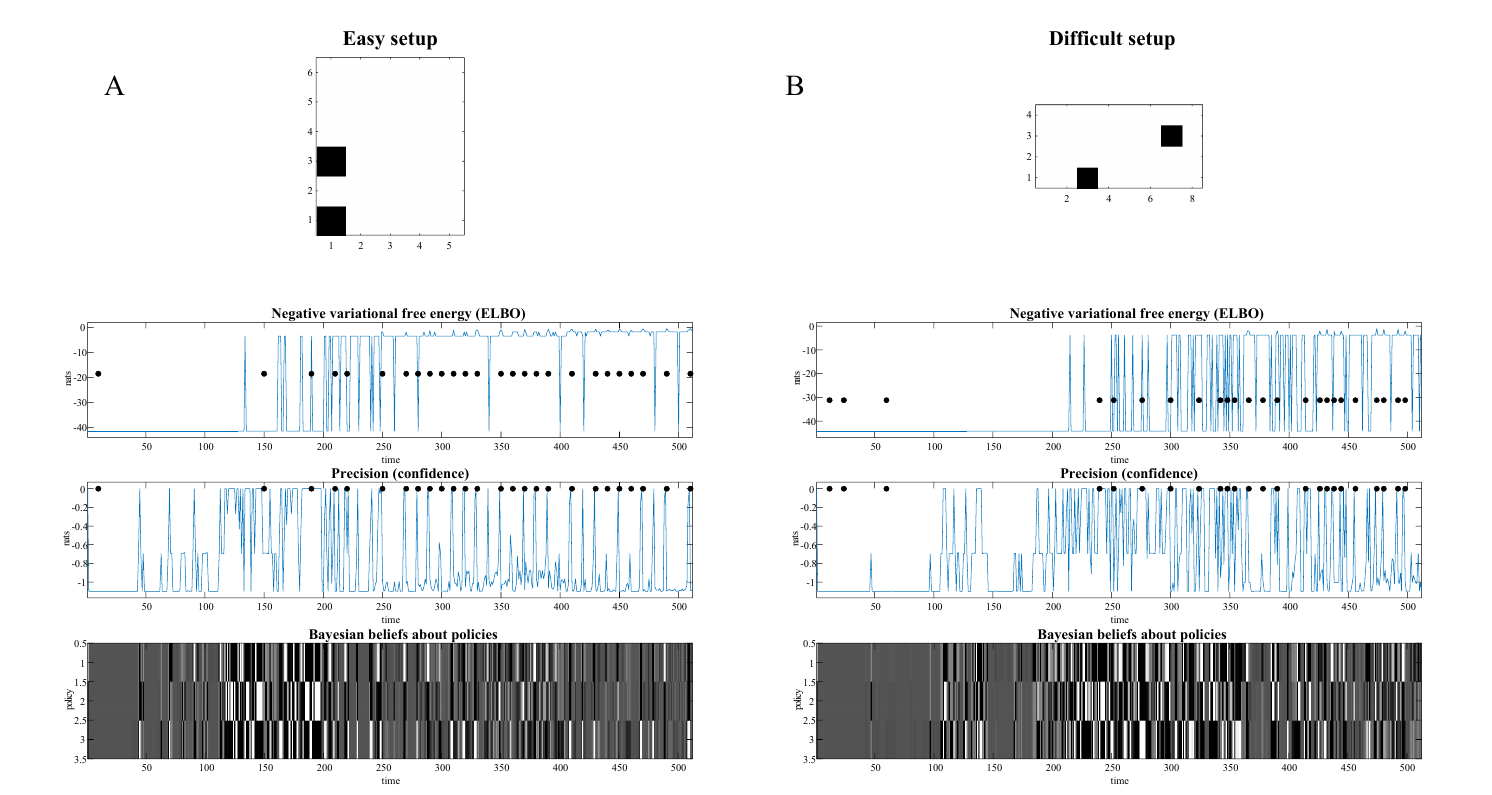}
    \caption{\textbf{The emergence of play.} Panels $A$ and $B$ show the results of two simulations of $512$ time steps (i.e., about two minutes of simulated time) under two configurations of the Pong set up: an easy setup in panel A and a slightly more difficult setup in panel B, in which the width of the bounding box was increased, and its height decreased (from $5 \times 6$ to $8 \times 4$). In both panels, the configuration of the game is shown above three plots reporting fluctuations in various measures of belief updating, and accompanying behaviour. The first graph plots the (negative) variational free energy as a function of time (where each time step corresponds roughly to $250$ ms). The black dots mark time points when the ball was hit. It can be seen that during accumulation of the likelihood Dirichlet counts, the ball was missed until time step $150$. After about a minute, the synthetic agent then starts to emit short rallies of between one and seven consecutive hits. The emergence of game play is accompanied by saltatory increases in negative variational free energy (or evidence lower bound). These increases disappear whenever the agent misses the ball, terminating little rallies. The second graph plots the average of the expected free energy under posterior beliefs about policies. This can be read as the precision of policy beliefs or, more colloquially, the confidence placed in policy selection. This illustrates that confident behaviour emerges during the first minute and is subsequently restricted to moments prior to hitting the ball. Heuristically, this can be read as the agent realising that it can avoid ambiguity by move moving in such a way as to catch the ball. The accompanying posterior (Bayesian) beliefs about policies are shown in image format in the lower plot. This illustrates that precise or confident behaviour entails precise beliefs about what to do next. Panel $B$ shows exactly the same results but for a slightly more difficult game. Here, the ball has more latitude to move horizontally and is returned more quickly, due to the reduced height of the bounding box. In consequence, learning a precise likelihood mapping takes about twice the amount of time. And, even when learned, the rallies are shorter, ranging from one to four, at most. We will use this more difficult set up to look at the effect of inductive planning in the next figure.}
\label{fig:pong_exp1}
\end{figure}

To simulate the \textit{in vitro} study, we exposed the synthetic neural network to $512$ observations—about two minutes of simulated time (i.e., a few seconds of computer time). Figure~\ref{fig:pong_exp1} shows the results of this simulation. The ensuing behaviour reproduced that observed empirically; namely, the emergence of short rallies after a minute or so of exposure. The question is now: can we understand this in terms of free energy minimising processes and their teleological concomitants?

As time progresses, Dirichlet counts are accumulated in the likelihood tensor to establish a precise mapping between each successive hidden state and the outcomes observed in each modality. This accumulation is precise because the agent has precise beliefs about state transitions. As the likelihood mapping is learned, it becomes apparent to the agent that certain states produce ambiguous outputs. These are the states in which it fails to hit the ball with the paddle. Because these ambiguous states have a high expected free energy—see Equation~\ref{eq:2}—the agent considers that actions that bring about these states are unlikely and therefore tries to avoid missing the ball. This is sufficient to support rallies of up to $7$ returns: see Figure~\ref{fig:pong_exp1}. 

However, because this agent does not look deep into the future, it can only elude ambiguous states when they are imminent. In other words, although this kind of behaviour can be regarded as sentient—in the sense that it rests upon an acquired model of the consequences of its own action—it is not equipped with intended states.

Note what has been simulated here does not rely on any notion of reinforcement learning: at no point was the agent rewarded for any behaviour or outcome. This kind of self-organisation—to a synchronous exchange with the world—is an emergent property of the system that simply rests on \textit{avoiding ambiguity or uncertainty} of a particular kind. The subtle distinction between a behaviourist (reinforcement learning) account and this kind of self-evidencing rests upon the imperatives for self-organised behaviour. In this \textit{in silico} reproduction of \textit{in vitro} experiments, behaviour is a consequence of (planning as) inference, where inference is based upon what has been learned. What has been learned are just statistical regularities (or unpredictable irregularities) in the environment: in this case, there are certain states that lead to unpredictable outcomes.  This gives the agent a precise grip on the world and enables it to infer its most likely actions. Its most likely actions are those that are characteristic of the thing it is; namely, something that minimises surprise, ambiguity, and free energy. This is distinct from learning a behaviour in the sense of reinforcement learning (e.g., a state-action mapping). The difference lies in the fact that behaviour—of the sort demonstrated above—rests on inference, under a learned model.

In the next section, we turn to a different kind of behaviour that rests upon inductive planning, equipping the agent with foresight and eliciting anticipatory behaviour. 

\subsection{Inductive Planning}

In this section, we repeat the simulations above, but making the game more difficult by increasing the width of the box. This means that to catch the ball, the agent has to anticipate outcomes in the distal future in order to respond with pre-emptive movement of the paddle. Note that this kind of behaviour goes beyond the sort of behaviour predicted under perceptual control theory and related accounts of ball catching \cite{gigerenzer2009homo,mansell2011control}. For example, one way to model behaviour in this paradigm would be to move the paddle so that it was always underneath the ball. However, this is not the behaviour that emerges under self-evidencing. In what follows, we will see that avoiding ambiguity is not sufficient for skilled performance of a more difficult game of Pong. However, if we equip the agent with intentions to hit the ball (i.e., as an intended state), it can use inductive planning to pursue a never ending rally, and play the game skilfully.

Figure~\ref{fig:pong_exp1}(B) reports performance over about two minutes of simulated time of an abductive agent when increasing the width of the Pong box to $8$ units (and decreasing its height to $4$ units). This simple change precludes sustained rallies; largely because the depth of planning is not sufficient to support pre-emptive moves of the paddle. 

\begin{figure}[H]
    \centering
    \includegraphics[width=0.85\textwidth]{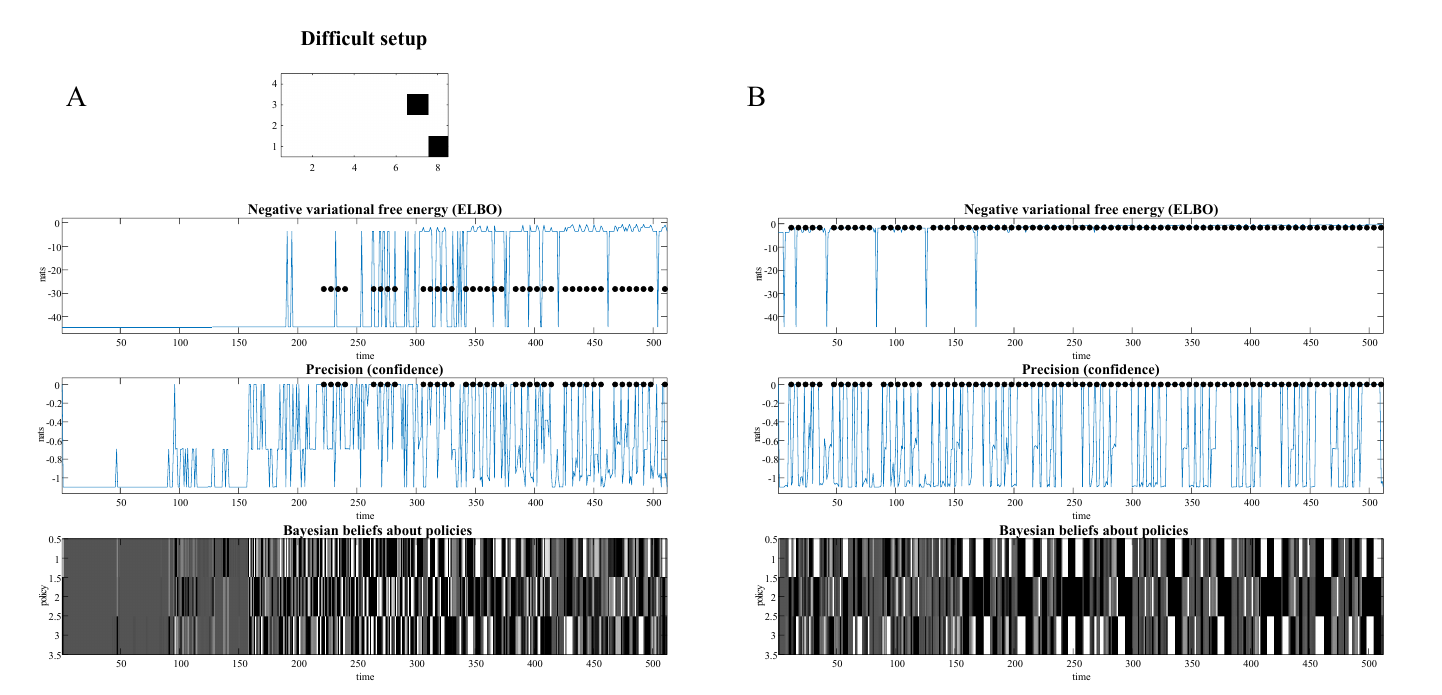}
    \caption{\textbf{Inductive planning.} This figure follows the same format as Figure 5, reporting the emergence of pong-playing behaviour under the more difficult set up described in the previous figure. However, here, we included inductive planning in the belief updating by specifying the agent's intentions in terms of priors over particular latent states; namely, states in which the agent hit the ball. In realising these intentions, the agent quickly learns a sufficiently precise likelihood mapping, evincing rallies of between four and six. after about a minute (of simulated time). This is shown in panel A. Panel B, shows the performance during the subsequent two minutes. By about three minutes, the agent has a precise grip on its world and realises its intentions fluently. From a dynamical systems perspective, this can be read as the emergence of generalised synchrony—or synchronisation of chaos—as the joint system converges onto a synchronisation manifold: a manifold that contains the states the agent intends to visit.}
\label{fig:inductive_exp}
\end{figure}

The equivalent results under inductive planning are shown in Figure~\ref{fig:inductive_exp}. Here, active inference under inductive constraints produces intermittent rallies within about a minute of simulated time—and skilled, and fluent play after three minutes.

In this example, we simply specified the intended states as those states corresponding to ball hits. This would be like instructing a child by telling her what is (i.e., which states are) expected of her. She can then work out how to realise those states by using inductive planning and selecting the most likely actions at each moment. Notice that there is no sense in which this could be construed as reinforcement learning: no reward or cost is being optimised, rather the behaviour is driven purely by the minimisation of uncertainty. A better metaphor would be instantiating some intentional set by instilling intentions or prior beliefs about characteristic states that should be realised. 

From the perspective of the free energy principle, priors over intended states can be cast as specifying a non-equilibrium steady-state with a (pullback) attractor that contains intended or characteristic states. From a dynamical systems perspective, this is equivalent to specifying unstable fixed points that characterise stable heteroclinic orbits \cite{afraimovich2008winnerless,rabinovich2008transient}, which have been discussed in terms of sequential behaviour \cite{fonollosa2015learning}. Intuitively, this means the agent has found a free energy minimum that is characterised by generalised synchrony between the neuronal network and the process generating sensory inputs. 

Given that this synchronisation was never seen in the \textit{in vitro} experiments, one might argue that the \emph{in vitro} behaviour was sentient but not intentional. In the remaining sections, we briefly showcase inductive planning in two other paradigms to illustrate the interaction between constraints—encoded by prior preferences over outcomes—and prior intentions, encoded by priors over latent states.

\section{Navigation as Inductive Planning}

In this section, we revisit a simple navigation problem addressed many times in the literature; e.g., \cite{baker2009action,dayan2006misbehavior}  and in demonstrations of active inference: e.g., \cite{friston2021sophisticated,kaplan2018planning}. Here, the problem is to learn the structure of a two-dimensional maze and then navigate to a target location based upon what has been learned. This features the dual problem of learning a world or generative model and then using what has been learned for deep planning and navigation.

\begin{figure}[H]
    \centering
    \includegraphics[width=0.9\textwidth]{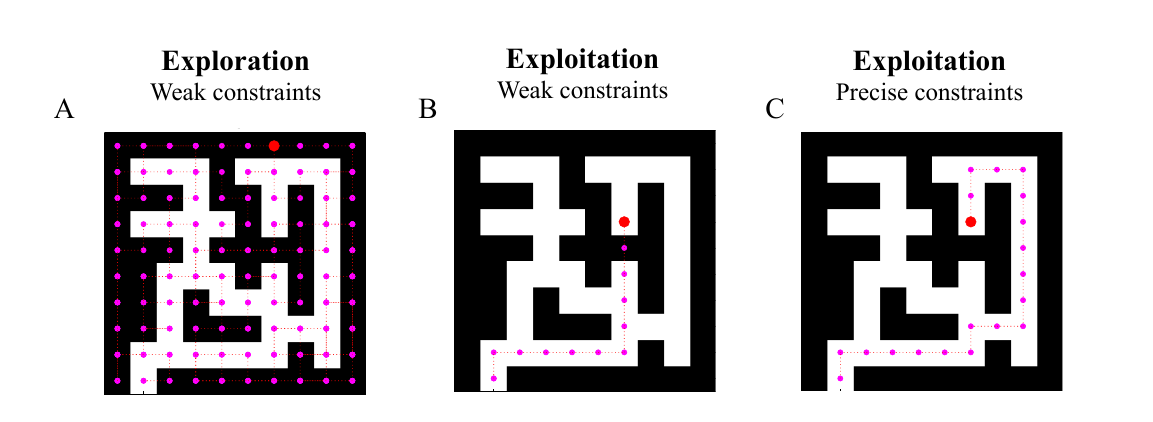}
    \caption{\textbf{Navigation by induction.} A: this panel reports the exploration of an agent that is building its likelihood mapping by exploring all the novel locations in a maze. Initially, the agent does not know where it can go; in the sense that it can only see its current location, which can be black or white. Therefore, every unvisited location furnishes some novelty; i.e., expected information gain (about likelihood parameters). This compels the agent to explore all locations efficiently and uniformly with an effective inhibition of return, until it has become familiar with this particular maze layout. After learning, the agent was given some intentions in terms of a specific location it believed, \textit{a priori}, it would visit. Panels B and C show the results of planning under mild and precise preferences for being on white squares. In panel B, the agent takes a short cut to the target location (red dot), which involves a transgression of one black square. This means that the cost of being on black squares is not sufficiently precise to have constrained the transitions used in inductive planning. However, because the agent is still trying to minimise expected cost (encoded by preferences for white squares) it navigates fairly gracefully until it encounters a barrier. In contrast, panel C shows the same agent with precise costs, which preclude transitions to black squares during inductive planning. This agent can swiftly induce the requisite path to the target location, without transgressing constraints on outcomes.}
\label{fig:navigation}
\end{figure}

In detail, we constructed a simple maze—shown in Figure~\ref{fig:navigation}—for an agent who has a myopic view of the world; namely, one output modality that reported whether the agent was sitting on an allowable location (white square) in the maze or a disallowed location (black square), which, \textit{a priori}, it found surprising (e.g., experiencing a foot shock). A simple generative model was supplied to the agent in the form of a single factor encoding each location or way-point, equipped with five paths. These were controllable paths that moved the agent up or down, or right or left (or staying still). The likelihood mapping was, as in the previous simulation, initialised to small uniform Dirichlet counts. This means the agent has no idea about the structure of its world but simply knew that a latent state could change in one of five ways. Learning this kind of environment is straightforward under active inference, due to the novelty or expected information gain about parameters (see Equation~\ref{eq:2}).

This means the agent chooses actions that resolve the greatest amount of uncertainty in the likelihood mapping from each latent state to outcomes. This ensures a Bayes optimal exploration of state space. 
Figure~\ref{fig:navigation}A shows that the agent pursues a path which covers all locations in an efficient fashion: i.e., not revisiting experienced states or locations until it has explored every location. The trajectory shown in Figure~\ref{fig:navigation}A corresponds to $256$ time steps. After this exposure, the agent has learned a likelihood model that is sufficient to support inductive planning. Figures~\ref{fig:navigation}B and C shows the results of this inductive navigation, reaching a distal target (red dot) from a starting location, while avoiding surprises or black squares in the maze. The two routes chosen are under imprecise and precise prior preferences for avoiding black squares (i.e., a log odds ratio—encoded by $\mathbf{c}$—of one and four, respectively). Note that the path under precise preferences is about $20$ steps, speaking to the depth of induction (here, $32$ time steps, as in Figure~\ref{fig:inductive}).

This example highlights an interesting aspect of inductive planning as defined here: namely, the learned constraints on foraging act as constraints on intentional behaviour. These constraints enter the allowable transitions, so that the paths that are induced respect the constraints due to prior preferences that can be inferred after—and only after—learning the likelihood mapping. In short, this example shows how it is possible to reach intended endpoints, under constraints on the way one gets there. In the final section, we use the same scheme to illustrate the efficiency of inductive planning in high dimensional problem spaces.

\section{Inductive Problem Solving}

This section considers a canonical problem solving task; namely, the Tower of Hanoi \cite{donnarumma2016problem}. In this problem, one has to rearrange a number of balls over a number of towers to reach a target arrangement from any given initial arrangement: see Figure 8. The problem can be made easier or more difficult by manipulating the number of intervening rearrangements between the initial and target (intended) configurations. We have previously shown that this problem can be learned from scratch using structure learning \cite{friston2023supervised}. Here, we consider problem-solving with and without inductive planning, after learning the likelihood model and allowable state transitions.

As above, implementing inductive planning simply means equipping the agent with prior beliefs about a final (intended) state and then letting it rearrange the balls until those intended states are realised. To solve this problem using active inference, one usually supplies constraints in terms of prior preferences that are mildly aversive for all but the target arrangement. This means the agent will rearrange the balls, in a state of mild surprise until the preferred arrangement is found—and the agent rests in a low free energy state. Because constraints are only in outcome space, there are certain arrangements that are less surprising because they are similar to the target configuration (as defined in outcome space). This enables the agent to solve fairly deep problems, even with a limited depth of planning (here, one-step-ahead planning). However, problems requiring more than four or five moves usually confound this kind of planning as inference. In contrast, if the intended target is specified in state space, then it will invoke inductive planning and, in principle, solve difficult problems, even with a limited depth of planning.

\begin{figure}[H]
    \centering
    \includegraphics[width=0.99\textwidth]{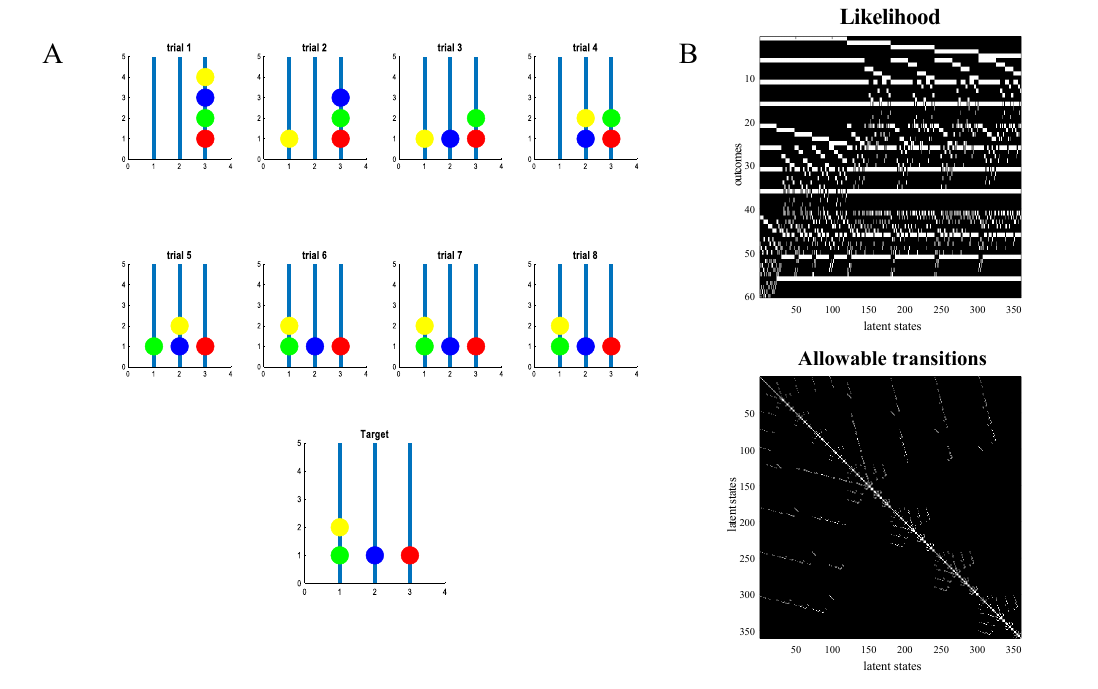}
    \caption{\textbf{Inductive planning and the Tower of Hanoi.} Panel A illustrates the particular game used to illustrate inductive planning. Here, there are four balls on three towers. The problem is to rearrange the initial configuration (on the upper left) to match the target configuration (lowest arrangement). In this example, it takes five moves. Actions correspond to moving a ball from one pillar to another. The generative model that supports this kind of problem solving is shown in terms of the requisite likelihood and transition mappings in panel B. The likelihood tensors have been stacked on top of each other (and unfolded) to illustrate the mapping between the $360$ latent states and the ($4 \times 3 \times 5$ =) $60$ outcomes. The accompanying transition parameters are shown in terms of allowable transitions among latent states (as in Figure~\ref{fig:inductive}). This generative model can be learned from scratch by presenting each arrangement—and then each rearrangement—of the balls to accumulate the appropriate Dirichlet parameters. Of interest here, is the use of the ensuing parameters or knowledge to solve problems that require deep planning. This problem is straightforward to solve using inductive planning; namely, working backwards from the target state using the protocol described in Figure~\ref{fig:inductive}. The ensuing performance is shown in the next figure.}
\label{fig:hanoi1}
\end{figure}

Figure~\ref{fig:hanoi2} shows the performance of two agents on $100$ problems, given $12$ moves for each problem. The first (abductive) agent was equipped only with constraints in outcome space; i.e., prior preferences that led to the target solution, provided that solution was reasonably close in outcome space. This agent failed to solve problems with five or more moves. In contrast, when specifying intentions in the form of the intended (target) state or arrangement, the second (inductive) agent was able to solve problems of eight moves or more almost instantaneously, without fail.

In these examples, the output space was a collection of $(4 \times 3 =) 12$ outcome modalities—one for each location or pixel—with five levels (four coloured balls or an empty outcome). The state space encompassed 360 arrangements, producing large ($360 \times 360 \times 5$) transition tensors. However, reducing these to logical matrices—used in inductive planning—means one can effectively plan deep into the future (here, $64$ moves) within milliseconds, using a one-step-ahead, active inference scheme.

\begin{figure}[H]
    \centering
    \includegraphics[width=0.7\textwidth]{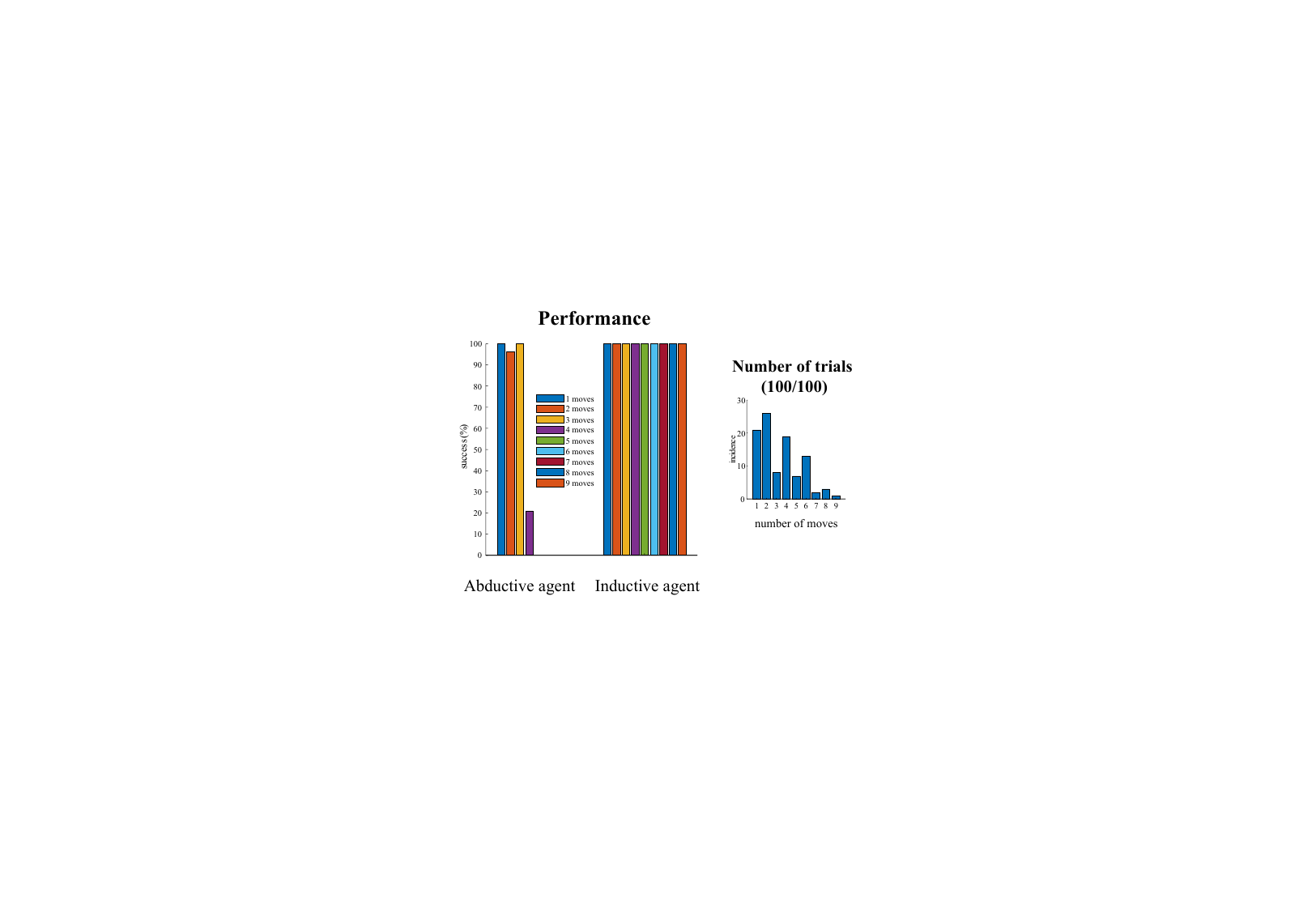}
    \caption{\textbf{Tower of Hanoi Performance.} This figure reports the performance of a generative model that has learned the Tower of Hanoi problem in terms of transitions among different arrangements of balls. We presented the agent with $100$ trials with different targets of greater and lesser difficulty (i.e., with varying numbers of moves from the initial and target arrangements). We presented exactly the same problems to agents with and without inductive planning. The right panel shows the incidence of trials in terms of the numbers of moves required until completion. The agent with inductive planning was able to solve $100\%$ of trials successfully. In contrast, the agent that did not use inductive planning was only able to complete problems of four moves or less. This is still impressive because both the abductive and inductive agents only looked one step ahead. In other words, even though the abductive agent could only evaluate the quality of its next move, it was still able to work towards the final solution. This is possible because the prior preferences for the target outcomes mean that certain outcomes are closer to the preferred outcomes than others. The 100 trials reported in this figure take less than 10 seconds to simulate.}
\label{fig:hanoi2}
\end{figure}

\section{Discussion}

This paper has introduced a particular instance of backwards induction to active inference, as well as a more formal characterisation of sentient and intentional behaviour. Induction in this setting appeals to a simple kind of backwards induction via logical operators, which is used to furnish constraints on the expected free energy, and hence, actions. Actions are then selected in the usual way; namely, actions that maximise expected information gain and value—where value is scored by log prior preferences over outcomes. The use of inductive priors lends planning a deep reach into the future that rests upon specifying final or intended endpoints. In turn, this differentiates sentient from intentional behaviour. To the extent that one can describe Bayesian beliefs—about the ultimate consequences of plans—as intentions, one could describe the behaviour illustrated above as intentional with a well-defined purpose or goal.

Inductive planning, as described here, can also be read as importing logical or symbolic (i.e. deductive) reasoning into a probabilistic (i.e., inductive, in the sense of inductive programming) framework. This speaks to symbolic approaches to problem solving and planning—e.g., \cite{colas2010common,fox2003pddl2,gilead2019above}—and a move towards the network tensor computations found in quantum computing: e.g., \cite{fields2023control,knill1997theory}. However, in so doing, one has to assume precise priors over state transitions and intended states. In other words, this kind of inductive planning is only apt when one has precisely stated goals and knowledge about state transitions. Is this a reasonable assumption for active inference? It could be argued that it is reasonable in the sense that: (i) goal-states or intended states are stipulatively precise (one cannot formulate an intention to act without specifying the intended outcome with a certain degree of precision) and (ii) the objective functions that underwrite self-evidencing lead to precise likelihood and transition mappings. In other words, to minimise expected free energy—via learning—just is to maximise the mutual information between latent states and their outcomes, and between successive latent states. 

To conclude, inductive planning differs from previous approaches proposed in both the reinforcement learning and active inference literature, due to the presence of intended goals  defined in latent state space. In both model free and model based reinforcement learning, goals are defined via a reward function. In alternative but similar approaches, such as active inference, rewards are passed to the agent as privileged (usually precise but sparse) observations \cite{friston2015active,dacosta2023reward}. This influences the behaviour of the agent, which learns to design and select policies that maximise expected future reward either via model-free approaches, which assign values to state-action pairs, or via model-based approaches, which select actions after simulating possible futures. Defining preferences directly in the state space, however, induces a different kind of behaviour: the fast and frugal computation involved in inductive planning is now apt to capture the efficiency of human-like decision-making, where indefinitely many possible paths, inconsistent with intended states, are ruled out \textit{a priori}—hence combining the ability of agents to seek long-term goals, with the efficiency of short-term planning.

\section{Conclusion}

The aim of this paper was to characterise the self-organisation of adaptive behaviour through the lens of the free energy principle, i.e., as self-evidencing. We did this by first discussing the definitions of reactive and sentient behaviour in active inference, where the latter describes the behaviour of agents that are aware of the consequences of their actions. We then introduced a formal account of \emph{intentional} behaviour, Specified by intended endpoints or goals, defined in state space rather than outcome space, as in abductive forms of active inference. We then investigate these forms of (reactive, sentient, and intentional) behaviour using simulations. First, we simulate the aforementioned \textit{in vitro} experiments, in which neuronal cultures spontaneously learn to play Pong, by implementing nested, free energy minimising processes. We used these simulations to Illustrate the ensuing  behaviour—leveraging the distinction between merely reactive, sentient, and intentional behaviour. The requisite inductive planning was then further illustrated using simple machine learning benchmarks (navigation in a grid world and the Tower of Hanoi problem), that showed how quickly and efficiently adaptive behaviour emerges under inductive constraints on active inference.

\section*{Disclosure statement}\label{sec:disclosure_statement}
The authors have no disclosures or conflict of interest.

\section*{Acknowledgements}
KF is supported by funding for the Wellcome Centre for Human Neuroimaging (Ref: 205103/Z/16/Z), a Canada-UK Artificial Intelligence Initiative (Ref: ES/T01279X/1) and the European Union’s Horizon 2020 Framework Programme for Research and Innovation under the Specific Grant Agreement No. 945539 (Human Brain Project SGA3). AR is funded by the Australian Research Council (Ref: DP200100757) and the Australian National Health and Medical Research Council (Investigator Grant Ref: 1194910).

\appendix
\include{universal_generative_models}

\begin{sloppypar}
\printbibliography[title={References}]

@article{kagan2022vitro,
  title={In vitro neurons learn and exhibit sentience when embodied in a simulated game-world},
  author={Kagan, B.J. and Kitchen, A.C. and Tran, N.T. and Habibollahi, F. and Khajehnejad, M. and Parker, B.J. and Bhat, A. and Rollo, B. and Razi, A. and Friston, K.J.},
  journal={Neuron},
  year={2022}
}

@article{todorov2006linearly,
  title={Linearly-solvable Markov decision problems},
  author={Todorov, Emanuel},
  journal={Advances in neural information processing systems},
  volume={19},
  year={2006}
}

@article{paul2023efficient,
  title={On efficient computation in active inference},
  author={Paul, Aswin and Sajid, Noor and Da Costa, Lancelot and Razi, Adeel},
  journal={arXiv preprint arXiv:2307.00504},
  year={2023}
}

@incollection{Kiefer2017-KIELPI,
	author = {Alex Kiefer},
	booktitle = {Philosophy and predictive processing},
	editor = {Thomas Metzinger and Wanja Wiese},
	title = {Literal Perceptual Inference},
	year = {2017}
}

@article{Smith2022-SMIAIM-4,
	author = {Ryan Smith and Maxwell J. D. Ramstead and Alex Kiefer},
	doi = {10.1007/s11229-022-03480-w},
	journal = {Synthese},
	number = {2},
	pages = {1--37},
	publisher = {Springer Verlag},
	title = {Active Inference Models Do Not Contradict Folk Psychology},
	volume = {200},
	year = {2022}
}

@article{fields2021minimal,
  title={Minimal physicalism as a scale-free substrate for cognition and consciousness},
  author={Fields, Chris and Glazebrook, James F and Levin, Michael},
  journal={Neuroscience of Consciousness},
  volume={2021},
  pages={niab013},
  year={2021}
}

@article{ghavamzadeh2015bayesian,
  title={Bayesian reinforcement learning: A survey},
  author={Ghavamzadeh, Mohammad and Mannor, Shie and Pineau, Joelle and Tamar, Aviv and others},
  journal={Foundations and Trends{\textregistered} in Machine Learning},
  volume={8},
  number={5-6},
  pages={359--483},
  year={2015},
  publisher={Now Publishers, Inc.}
}

@article{knill1997theory,
  title={Theory of quantum error-correcting codes},
  author={Knill, E. and Laflamme, R.},
  journal={Physical Review A},
  volume={55},
  pages={900--911},
  year={1997}
}

@article{botvinick2012planning,
  title={Planning as inference},
  author={Botvinick, Matthew and Toussaint, Marc},
  journal={Trends in cognitive sciences},
  volume={16},
  number={10},
  pages={485--488},
  year={2012},
  publisher={Elsevier}
}

@book{berger2013statistical,
  title={Statistical decision theory and Bayesian analysis},
  author={Berger, James O},
  year={2013},
  publisher={Springer Science \& Business Media}
}

@article{friston2021sophisticated,
  title={Sophisticated inference},
  author={Friston, K. and Da Costa, L. and Hafner, D. and Hesp, C. and Parr, T.},
  journal={Neural Computation},
  volume={33},
  pages={713--763},
  year={2021}
}

@article{howard1966information,
  title={Information value theory},
  author={Howard, Ronald A},
  journal={IEEE Transactions on systems science and cybernetics},
  volume={2},
  number={1},
  pages={22--26},
  year={1966},
  publisher={IEEE}
}

@article{winn2005variational,
  title={Variational message passing},
  author={Winn, J. and Bishop, C.M.},
  journal={Journal of Machine Learning Research},
  volume={6},
  pages={661--694},
  year={2005}
}

@inproceedings{braun2011path,
  title={Path integral control and bounded rationality},
  author={Braun, Daniel A and Ortega, Pedro A and Theodorou, Evangelos and Schaal, Stefan},
  booktitle={2011 IEEE symposium on adaptive dynamic programming and reinforcement learning (ADPRL)},
  pages={202--209},
  year={2011},
  organization={IEEE}
}

@article{dacosta2020active,
  title={Active inference on discrete state-spaces: A synthesis},
  author={Da Costa, Lancelot and Parr, Thomas and Sajid, Noor and Veselic, Sanja and Neacsu, Victor and Friston, Karl},
  journal={Journal of Mathematical Psychology},
  volume={99},
  pages={102447},
  year={2020}
}

@article{ramstead2023bayesian,
  title={On Bayesian mechanics: a physics of and by beliefs},
  author={Ramstead, Maxwell JD and Sakthivadivel, Dalton AR and Heins, Conor and Koudahl, Magnus and Millidge, Beren and Da Costa, Lancelot and Klein, Brennan and Friston, Karl J},
  journal={Interface Focus},
  volume={13},
  number={3},
  pages={20220029},
  year={2023},
  publisher={The Royal Society}
}

@article{hure2020deep,
  title={Deep Backward Schemes for High-Dimensional Nonlinear Pdes},
  author={Hure, C. and Pham, H. and Warin, X.},
  journal={Mathematics of Computation},
  volume={89},
  pages={1547--1579},
  year={2020}
}

@article{still2012information,
  title={An information-theoretic approach to curiosity-driven reinforcement learning},
  author={Still, S. and Precup, D.},
  journal={Theory in Biosciences},
  volume={131},
  pages={139--148},
  year={2012}
}

@misc{friston2022path,
  title={Path integrals, particular kinds, and strange things},
  author={Friston, K. and Da Costa, L. and Sakthivadivel, D.A.R. and Heins, C. and Pavliotis, G.A. and Ramstead, M. and Parr, T.},
  note={arXiv:2210.12761},
  year={2022}
}

@inproceedings{attias2003planning,
  title={Planning by probabilistic inference},
  author={Attias, Hagai},
  booktitle={International workshop on artificial intelligence and statistics},
  pages={9--16},
  year={2003},
  organization={PMLR}
}

@article{itti2009bayesian,
  title={Bayesian Surprise Attracts Human Attention},
  author={Itti, L. and Baldi, P.},
  journal={Vision Research},
  volume={49},
  pages={1295--1306},
  year={2009}
}

@article{sandved-smith2021towards,
  title={Towards a computational phenomenology of mental action: modelling meta-awareness and attentional control with deep parametric active inference},
  author={Sandved-Smith, L. and Hesp, C. and Mattout, J. and Friston, K. and Lutz, A. and Ramstead, M.J.D.},
  journal={Neuroscience of Consciousness},
  volume={2021},
  pages={niab018},
  year={2021}
}

@phdthesis{beal2003variational,
  title={Variational Algorithms for Approximate Bayesian Inference},
  author={Beal, Matthew J},
  year={2003},
  school={University College London}
}

@incollection{seth2015inference,
  title={Inference to the Best Prediction},
  author={Seth, A.K.},
  booktitle={Open MIND},
  editor={Metzinger, T.K. and Windt, J.M.},
  publisher={MIND Group},
  address={Frankfurt am Main},
  year={2015}
}

@article{schwartenbeck2019computational,
  title={Computational mechanisms of curiosity and goal-directed exploration},
  author={Schwartenbeck, Philipp and Passecker, Johannes and Hauser, Tobias U and FitzGerald, Thomas HB and Kronbichler, Martin and Friston, Karl J},
  journal={elife},
  volume={8},
  pages={e41703},
  year={2019},
  publisher={eLife Sciences Publications, Ltd}
}

@article{tervo2016toward,
  title={Toward the neural implementation of structure learning},
  author={Tervo, D Gowanlock R and Tenenbaum, Joshua B and Gershman, Samuel J},
  journal={Current opinion in neurobiology},
  volume={37},
  pages={99--105},
  year={2016},
  publisher={Elsevier}
}

@article{olshausen1996emergence,
  title={Emergence of simple-cell receptive field properties by learning a sparse code for natural images},
  author={Olshausen, B.A. and Field, D.J.},
  journal={Nature},
  volume={381},
  pages={607--609},
  year={1996}
}

@article{isomura2018vitro,
  title={In vitro neural networks minimise variational free energy},
  author={Isomura, T. and Friston, K.},
  journal={Scientific Reports},
  volume={8},
  pages={16926},
  year={2018}
}

@article{barlow1961possible,
  title={Possible principles underlying the transformation of sensory messages},
  author={Barlow, Horace B and others},
  journal={Sensory communication},
  volume={1},
  number={01},
  pages={217--233},
  year={1961}
}

@inproceedings{sanchez2020learning,
  title={Learning disentangled representations via mutual information estimation},
  author={Sanchez, Eduardo Hugo and Serrurier, Mathieu and Ortner, Mathias},
  booktitle={Computer Vision--ECCV 2020: 16th European Conference, Glasgow, UK, August 23--28, 2020, Proceedings, Part XXII 16},
  pages={205--221},
  year={2020},
  organization={Springer}
}

@article{broek2012risk,
  title={Risk sensitive path integral control},
  author={Broek, Bart van den and Wiegerinck, Wim and Kappen, Hilbert},
  journal={arXiv preprint arXiv:1203.3523},
  year={2012}
}

@article{fields2023control,
  title={Control flow in active inference systems—part II: tensor networks as general models of control flow},
  author={Fields, Chris and others},
  journal={IEEE Transactions on Molecular, Biological and Multi-Scale Communications},
  volume={9},
  pages={246--256},
  year={2023}
}

@article{van2013informational,
  title={Informational constraints-driven organization in goal-directed behavior},
  author={Van Dijk, Sander G and Polani, Daniel},
  journal={Advances in Complex Systems},
  volume={16},
  number={02n03},
  pages={1350016},
  year={2013},
  publisher={World Scientific}
}

@misc{schrittwieser2019mastering,
  title={Mastering Atari, Go, Chess and Shogi by Planning with a Learned Model},
  author={Schrittwieser, J. and Antonoglou, I. and Hubert, T. and Simonyan, K. and Sifre, L. and Schmitt, S. and Guez, A. and Lockhart, E. and Hassabis, D. and Graepel, T. and Lillicrap, T.P. and Silver, D.},
  note={arXiv:1911.08265},
  year={2019}
}

@article{friston2015active,
  title={Active inference and epistemic value},
  author={Friston, Karl and Rigoli, Francesco and Ognibene, Dimitri and Mathys, Christoph and Fitzgerald, Thomas and Pezzulo, Giovanni},
  journal={Cognitive neuroscience},
  volume={6},
  number={4},
  pages={187--214},
  year={2015},
  publisher={Taylor \& Francis}
}

@article{barlow1974inductive,
  title={Inductive inference, coding, perception, and language},
  author={Barlow, Horace B},
  journal={Perception},
  volume={3},
  pages={123--134},
  year={1974}
}

@inproceedings{schmidhuber1991curious,
  title={Curious model-building control systems},
  author={Schmidhuber, J.},
  booktitle={International Joint Conference on Neural Networks},
  volume={2},
  pages={1458--1463},
  organization={IEEE},
  year={1991}
}

@article{kaplan2018planning,
  title={Planning and navigation as active inference},
  author={Kaplan, R. and Friston, K.J.},
  journal={Biological Cybernetics},
  volume={112},
  pages={323--343},
  year={2018}
}

@article{ay2008predictive,
  title={Predictive information and explorative behavior of autonomous robots},
  author={Ay, Nihat and Bertschinger, Nils and Der, Ralf and Guttler, Frank and Olbrich, Eckehard},
  journal={European Physical Journal B},
  volume={63},
  pages={329--339},
  year={2008}
}

@article{lindley1956measure,
  title={On a Measure of the Information Provided by an Experiment},
  author={Lindley, D.V.},
  journal={Annals of Mathematical Statistics},
  volume={27},
  pages={986--1005},
  year={1956}
}

@article{bellman1952dynamic,
  title={On the Theory of Dynamic Programming},
  author={Bellman, Richard},
  journal={Proc Natl Acad Sci U S A},
  volume={38},
  pages={716--719},
  year={1952}
}

@misc{ramstead2022bayesian,
  title={On Bayesian Mechanics: A Physics of and by Beliefs},
  author={Ramstead, M.J.D. and Sakthivadivel, D.A.R. and Heins, C. and Koudahl, M. and Millidge, B. and Da Costa, L. and Klein, B. and Friston, K.J.},
  note={arXiv:2205.11543},
  year={2022}
}

@article{rabinovich2008transient,
  title={Transient dynamics for neural processing},
  author={Rabinovich, M. and Huerta, R. and Laurent, G.},
  journal={Science},
  volume={321},
  pages={48--50},
  year={2008}
}

@article{baker2009action,
  title={Action understanding as inverse planning},
  author={Baker, Chris L and Saxe, Rebecca and Tenenbaum, Joshua B},
  journal={Cognition},
  volume={113},
  pages={329--349},
  year={2009}
}

@article{clark2019bayesing,
  title={Bayesing Qualia: Consciousness as Inference, Not Raw Datum},
  author={Clark, Andy and Friston, Karl and Wilkinson, Sam},
  journal={Journal of Consciousness Studies},
  volume={26},
  pages={19--33},
  year={2019}
}

@article{colas2010common,
  title={Common Bayesian models for common cognitive issues},
  author={Colas, Fr{\'e}d{\'e}ric and Diard, Julien and Bessi{\`e}re, Pierre},
  journal={Acta Biotheoretica},
  volume={58},
  pages={191--216},
  year={2010}
}

@article{higgins2021unsupervised,
  title={Unsupervised deep learning identifies semantic disentanglement in single inferotemporal face patch neurons},
  author={Higgins, Irina and Chang, Le and Langston, Victoria and Hassabis, Demis and Summerfield, Christopher and Tsao, Doris and Botvinick, Matthew},
  journal={Nature communications},
  volume={12},
  number={1},
  pages={6456},
  year={2021},
  publisher={Nature Publishing Group UK London}
}

@article{camerer2004cognitive,
  title={A cognitive hierarchy model of games},
  author={Camerer, Colin F and Ho, Teck-Hua and Chong, Juin-Kuan},
  journal={Quarterly Journal of Economics},
  volume={119},
  pages={861--898},
  year={2004}
}

@article{mansell2011control,
  title={Control of perception should be operationalized as a fundamental property of the nervous system},
  author={Mansell, W.},
  journal={Topics in Cognitive Science},
  volume={3},
  pages={257--261},
  year={2011}
}

@article{sakthivadivel2022weak,
  title={Weak Markov blankets in high-dimensional, sparsely-coupled random dynamical systems},
  author={Sakthivadivel, Dalton AR},
  journal={arXiv preprint arXiv:2207.07620},
  year={2022}
}

@article{mnih2015human,
  title={Human-level control through deep reinforcement learning},
  author={Mnih, V. and Kavukcuoglu, K. and Silver, D. and Rusu, A.A. and Veness, J. and Bellemare, M.G. and Graves, A. and Riedmiller, M. and Fidjeland, A.K. and Ostrovski, G. and Petersen, S. and Beattie, C. and Sadik, A. and Antonoglou, I. and King, H. and Kumaran, D. and Wierstra, D. and Legg, S. and Hassabis, D.},
  journal={Nature},
  volume={518},
  pages={529--533},
  year={2015}
}

@article{linsker1990perceptual,
  title={Perceptual Neural Organization - Some Approaches Based on Network Models and Information-Theory},
  author={Linsker, R.},
  journal={Annual Review of Neuroscience},
  volume={13},
  pages={257--281},
  year={1990}
}

@article{gilead2019above,
  title={Above and beyond the concrete: The diverse representational substrates of the predictive brain},
  author={Gilead, M. and Trope, Y. and Liberman, N.},
  journal={Behavioral and Brain Sciences},
  volume={43},
  pages={e121},
  year={2019}
}

@article{pezzulo2014internally,
  title={Internally generated sequences in learning and executing goal-directed behavior},
  author={Pezzulo, Giovanni and Van der Meer, Matthijs AA and Lansink, Carien S and Pennartz, Cyriel MA},
  journal={Trends in cognitive sciences},
  volume={18},
  number={12},
  pages={647--657},
  year={2014},
  publisher={Elsevier}
}

@article{penny2013forward,
  title={Forward and backward inference in spatial cognition},
  author={Penny, Will D and Zeidman, Peter and Burgess, Neil},
  journal={PLoS computational biology},
  volume={9},
  number={12},
  pages={e1003383},
  year={2013},
  publisher={Public Library of Science San Francisco, USA}
}

@article{louie2001temporally,
  title={Temporally structured replay of awake hippocampal ensemble activity during rapid eye movement sleep},
  author={Louie, Kenway and Wilson, Matthew A},
  journal={Neuron},
  volume={29},
  number={1},
  pages={145--156},
  year={2001},
  publisher={Elsevier}
}

@article{sutton1999between,
  title={Between MDPs and semi-MDPs: A framework for temporal abstraction in reinforcement learning},
  author={Sutton, R.S. and Precup, D. and Singh, S.},
  journal={Artificial Intelligence},
  volume={112},
  pages={181--211},
  year={1999}
}

@article{dacosta2023reward,
  title={Reward Maximization Through Discrete Active Inference},
  author={Da Costa, Lancelot and Sajid, Noor and Parr, Thomas and Friston, Karl and Smith, Ryan},
  journal={Neural Computation},
  volume={35},
  number={5},
  pages={807--852},
  year={2023},
  publisher={MIT Press}
}

@article{friston2023supervised,
  title={Supervised structure learning},
  author={Friston, Karl and Da Costa, Lancelot and Tschantz, Alexander and Kiefer, Alex and Salvatori, Tommaso and Neacsu, Victorita, and Koudahl, Magnus and Heins, Conor and Sajid, Noor and Markovic, Dimitrije and Parr, Thomas and Verbelen, Tim and Buckley, Christopher},
  journal={arXiv preprint arXiv:2311.10300},
  year={2023}
}

@article{dacosta2020relationship,
  title={The relationship between dynamic programming and active inference: the discrete, finite-horizon case},
  author={Da Costa, Lancelot and Sajid, Noor and Parr, Thomas and Friston, Karl and Smith, Ryan},
  journal={arXiv preprint arXiv:2009.08111},
  year={2020}
}

@article{dayan2006misbehavior,
  title={The misbehavior of value and the discipline of the will},
  author={Dayan, Peter and Niv, Yael and Seymour, Ben and Daw, Nathaniel D},
  journal={Neural Networks},
  volume={19},
  pages={1153--1160},
  year={2006}
}

@book{howard1960dynamic,
  title={Dynamic Programming and Markov Processes},
  author={Howard, R.A.},
  publisher={MIT Press},
  address={Cambridge, MA},
  year={1960}
}

@misc{hawthorne2021inductive,
  title={Inductive Logic},
  author={Hawthorne, J.},
  note={Stanford Encyclopedia of Philosophy},
  year={2021}
}

@article{palacios2019emergence,
  title={The emergence of synchrony in networks of mutually inferring neurons},
  author={Palacios, E.R. and Isomura, T. and Parr, T. and Friston, K.},
  journal={Scientific Reports},
  volume={9},
  pages={6412},
  year={2019}
}

@book{parr2022active,
  title={Active inference: the free energy principle in mind, brain, and behavior},
  author={Parr, Thomas and Pezzulo, Giovanni and Friston, Karl J},
  year={2022},
  publisher={MIT Press}
}

@article{fox2003pddl2,
  title={PDDL2.1: An extension to PDDL for expressing temporal planning domains},
  author={Fox, Maria and Long, Derek},
  journal={Journal of Artificial Intelligence Research},
  volume={20},
  pages={61--124},
  year={2003}
}

@article{balci2023response,
  title={A response to claims of emergent intelligence and sentience in a dish},
  author={Balci, Fuat and others},
  journal={Neuron},
  volume={111},
  pages={604--605},
  year={2023}
}

@article{afraimovich2008winnerless,
  title={Winnerless competition principle and prediction of the transient dynamics in a Lotka-Volterra model},
  author={Afraimovich, Valentin and Tristan, Irma and Huerta, Ram{\'o}n and Rabinovich, Mikhail I},
  journal={Chaos},
  volume={18},
  number={043103},
  year={2008}
}

@article{levin2019computational,
  title={The Computational Boundary of a "Self": Developmental Bioelectricity Drives Multicellularity and Scale-Free Cognition},
  author={Levin, M.},
  journal={Frontiers in Psychology},
  volume={10},
  pages={2688},
  year={2019}
}

@inproceedings{klyubin2005empowerment,
  title={Empowerment: A universal agent-centric measure of control},
  author={Klyubin, Alexander S and Polani, Daniel and Nehaniv, Chrystopher L},
  booktitle={2005 ieee congress on evolutionary computation},
  volume={1},
  pages={128--135},
  year={2005},
  organization={IEEE}
}

@article{tipping2001sparse,
  title={Sparse Bayesian learning and the relevance vector machine},
  author={Tipping, Michael E},
  journal={Journal of machine learning research},
  volume={1},
  number={Jun},
  pages={211--244},
  year={2001}
}

@article{camerer1997progress,
  title={Progress in behavioral game theory},
  author={Camerer, Colin F},
  journal={Journal of Economic Perspectives},
  volume={11},
  pages={167--188},
  year={1997}
}

@article{gigerenzer2009homo,
  title={Homo heuristicus: why biased minds make better inferences},
  author={Gigerenzer, G. and Brighton, H.},
  journal={Topics in Cognitive Science},
  volume={1},
  pages={107--143},
  year={2009}
}

@article{hohwy2016self,
  title={The self-evidencing brain},
  author={Hohwy, Jakob},
  journal={No{\^u}s},
  volume={50},
  number={2},
  pages={259--285},
  year={2016},
  publisher={Wiley Online Library}
}

@article{buckner2010role,
  title={The role of the hippocampus in prediction and imagination},
  author={Buckner, Randy L},
  journal={Annual review of psychology},
  volume={61},
  pages={27--48},
  year={2010},
  publisher={Annual Reviews}
}

@article{isomura2023experimental,
  title={Experimental validation of the free-energy principle with in vitro neural networks},
  author={Isomura, T. and Kotani, K. and Jimbo, Y. and Friston, K.J.},
  journal={Nature Communications},
  volume={14},
  pages={4547},
  year={2023}
}

@article{donnarumma2016problem,
  title={Problem solving as probabilistic inference with subgoaling: explaining human successes and pitfalls in the tower of hanoi},
  author={Donnarumma, Francesco and Maisto, Domenico and Pezzulo, Giovanni},
  journal={PLoS computational biology},
  volume={12},
  number={4},
  pages={e1004864},
  year={2016},
  publisher={Public Library of Science San Francisco, CA USA}
}

@article{gros2009cognitive,
  title={Cognitive computation with autonomously active neural networks: An emerging field},
  author={Gros, C.},
  journal={Cognitive Computation},
  volume={1},
  pages={77--90},
  year={2009}
}

@article{fonollosa2015learning,
  title={Learning of chunking sequences in cognition and behavior},
  author={Fonollosa, Jos{\'e} and Neftci, Emre and Rabinovich, Mikhail},
  journal={PLoS Computational Biology},
  volume={11},
  pages={e1004592},
  year={2015}
}

@misc{manicka2019modeling,
  title={Modeling somatic computation with non-neural bioelectric networks},
  author={Manicka, S. and Levin, M.},
  note={Scientific Reports 9, 18612},
  year={2019}
}

@article{friston2023free,
  title={The free energy principle made simpler but not too simple},
  author={Friston, K. and Da Costa, L. and Sajid, N. and Heins, C. and Ueltzh{\"o}ffer, K. and Pavliotis, G.A. and Parr, T.},
  journal={Physics Reports},
  volume={1024},
  pages={1--29},
  year={2023}
}

@article{ortega2013thermodynamics,
  title={Thermodynamics as a theory of decision-making with information-processing costs},
  author={Ortega, Pedro A and Braun, Daniel A},
  journal={Proceedings of the Royal Society A: Mathematical, Physical and Engineering Sciences},
  volume={469},
  number={2153},
  pages={20120683},
  year={2013},
  publisher={The Royal Society Publishing}
}

@article{harman1965inference,
  title={The inference to the best explanation},
  author={Harman, G.H.},
  journal={Philosophical Review},
  volume={74},
  pages={88--95},
  year={1965}
}

@inproceedings{masumori2015emergence,
  title={Emergence of sense-making behavior by the Stimulus Avoidance Principle: Experiments on a robot behavior controlled by cultured neuronal cells},
  author={Masumori, Atsushi and Maruyama, Norihiro and Sinapayen, Lana and Mita, Takeshi and Frey, Urs and Bakkum, Douglas and Takahashi, Hirokazu and Ikegami, Takashi},
  booktitle={Artificial Life Conference Proceedings},
  pages={373--380},
  year={2015},
  organization={MIT Press One Rogers Street, Cambridge, MA 02142-1209, USA journals-info~…}
}
\end{sloppypar}

\end{document}